\documentclass[journal]{IEEEtran}
\usepackage{amsthm}
\usepackage{empheq}
\usepackage{cancel}
\usepackage{mdframed,xcolor}
\usepackage{tcolorbox}
\usepackage{placeins}
\usepackage{relsize}
\usepackage{setspace}
\usepackage{stmaryrd}
\usepackage{xcolor}
\usepackage{url}
\usepackage{multirow}
\usepackage{amssymb}
\usepackage{amsmath}
\usepackage[end]{algpseudocode}
\usepackage{algorithm}
\usepackage{amsfonts}
\usepackage{diagbox}
\usepackage{booktabs}
\usepackage{filecontents}
\usepackage{mathtools}
\usepackage{caption}
\usepackage{subcaption}
\usepackage{enumitem}
\usepackage{xfrac}
\usepackage{nicefrac}
\usepackage{soul}
\usepackage{algorithm}
\algdef{SE}{Begin}{End}{\textbf{begin}}{\textbf{end}}
\setlist[itemize]{leftmargin=*}
\usepackage{xcolor}
\definecolor{rv1}{rgb}{1.0, 0.44, 0.37}
\definecolor{rv2}{rgb}{0.4, 1.0, 0.0}
\definecolor{rv3}{rgb}{0.0, 0.75, 1.0}
\definecolor{rvt}{rgb}{0.75, 0.75, 0.75}
\definecolor{my-blue}{cmyk}{0.1, 0.0, 0.0, 0.0, 1.00}
\usepackage{soul}

\newtheoremstyle{exampstyle}
{7pt} 
{7pt} 
{\itshape} 
{} 
{\bfseries} 
{.} 
{.5em} 
{} 
\theoremstyle{exampstyle}

\newcommand{\algrule}[1][.2pt]{\par\vskip.5\baselineskip\hrule height #1\par\vskip.5\baselineskip}

\newsavebox\mybox

\begin{document}

\title{Energy Efficient Design of Active STAR-RIS-Aided SWIPT Systems}

\author{Sajad Faramarzi, Hosein Zarini, Sepideh Javadi, Mohammad Robat Mili, Rui Zhang, \textit{Fellow}, \textit{IEEE}, \\George K. Karagiannidis, \textit{Fellow}, \textit{IEEE}, and Naofal Al-Dhahir, \textit{Fellow}, \textit{IEEE}
	\thanks{Sajad Faramarzi is with the School of Electrical Engineering, Iran University of Science $\&$ Technology, Tehran, Iran (e-mail: sajad.faramarzi$1397$@gmail.com) \newline Hosein Zarini is with the Dept. of Computer Engineering, Sharif University of Technology, Tehran, Iran (e-mail: hosein.zarini68@sharif.edu)\newline Sepideh Javadi,  and Mohammad Robat Mili are with the Pasargad Institute for Advanced Innovative Solutions (PIAIS), Tehran, Iran (e-mail: $\lbrace$sepideh.javadi, mohammad.robatmili$\rbrace$@piais.ir)\newline R. Zhang is with School of Science and Engineering, Shenzhen Research Institute of Big Data, The Chinese University of Hong Kong, Shenzhen, Guangdong 518172, China. He is also with the Department of Electrical and Computer Engineering, National University of Singapore, Singapore 117583 (e-mail: elezhang@nus.edu.sg)\newline George K. Karagiannidis is with Department of Electrical and Computer
    Engineering, Aristotle University of Thessaloniki, Greece and also with Artificial Intelligence $\&$ Cyber Systems Research Center, Lebanese American
    University (LAU), Lebanon (e-mail: geokarag@auth.gr)\newline Naofal Al-Dhahir is with the Department of Electrical and Computer Engineering, The University of Texas at Dallas, Richardson, TX 75080 USA (aldhahir@utdallas.edu). }}

\markboth{}%
{Shell \MakeLowercase{\textit{et al.}}: Bare Demo of IEEEtran.cls for IEEE Journals}

\maketitle
\begin{abstract}
	In this paper, we consider the downlink transmission of a multi-antenna base station (BS) supported by an active simultaneously transmitting and reconfigurable intelligent surface (STAR-RIS) to serve single-antenna users via simultaneous wireless information and power transfer (SWIPT). In this context, we formulate an energy efficiency maximisation problem that jointly optimises the gain, element selection and phase shift matrices of the active STAR-RIS, the transmit beamforming of the BS and the power splitting ratio of the users. With respect to the highly coupled and non-convex form of this problem, an alternating optimisation solution approach is proposed, using tools from convex optimisation and reinforcement learning. Specifically, semi-definite relaxation (SDR), difference of concave functions (DC), and fractional programming techniques are employed to transform the non-convex optimisation problem into a convex form for optimising the BS beamforming vector and the power splitting ratio of the SWIPT. Then, by integrating meta-learning with the modified deep deterministic policy gradient (DDPG) and soft actor-critical (SAC) methods, a combinatorial reinforcement learning network is developed to optimise the element selection, gain and phase shift matrices of the active STAR-RIS. Our simulations show the effectiveness of the proposed resource allocation scheme. Furthermore, our proposed active STAR-RIS-based SWIPT system outperforms its passive counterpart by $57\%$ on average.
\end{abstract}
\vspace*{-0.1em}
\vspace*{-0.1em}
\IEEEpeerreviewmaketitle
\vspace{-1.05em}
\section{Introduction}
\subsection{Background and Motivation}
In recent years, the emergence of Internet of Things (IoT) devices, such as smartphones, monitoring sensors and wearable devices, with limited energy storage imposes a stringent requirement on energy harvesting (EH) in sixth generation (6G) wireless networks~\cite{IoT-EH}. 
Simultaneous Wireless Information and Power Transfer (SWIPT) technology, by superimposing information and energy within the power domain of a unified signal, has recently been viewed as a potential solution ~\cite{es2}. In SWIPT systems, the receiver is equipped with a power splitter (PS) for information
decoding (ID) and EH at the same time. Thanks to the EH mechanism, the power constraints of IoT devices can be alleviated without any appreciable loss of performance in terms of communication rate/range~\cite{es3},~\cite{J4}.
\textcolor{black}{Another barrier to the development of IoT devices is their massive scalability. To support a large number of ubiquitous IoT devices distributed everywhere, it is essential to extend the accessibility and coverage of existing wireless networks. In this context, a reconfigurable intelligent surface (RIS) is able to bypass the environmental obstacles and initiate virtual line-of-sight (LoS) links~\cite{Hosein4}, which remarkably expands the coverage area and thus the number of supported IoT devices~\cite{Outage, SLIPT}.} A RIS architecturally incorporates passive metasurfaces capable of reflecting incoming signals without relying on a dedicated power source~\cite{Hosein5}. Due to the multiplicative fading effect, conventional passive beamforming RISs suffer from severe signal strength attenuation and possible detection problems at the receiver~\cite{es4}. Active RIS, on the other hand, overcome this disadvantage and enjoy strong transmitted signals at the expense of higher energy consumption due to an amplification mechanism~\cite{es7}. \textcolor{black}{In \cite{ZhuHan}, the authors proposed an energy efficient element selection mechanism to optimise the on/off state of the RIS elements, which strongly controls the power consumption, especially in active RISs. In \cite{EE}, it was numerically demonstrated that an efficient element selection mechanism remarkably promotes the energy efficiency (EE) of a RIS-supported network, and for larger RISs this gain is even more pronounced.}
Recently, the Simultaneous Transmitting and Reflecting Reconfigurable Intelligent Surface (STAR-RIS)~\cite{STAR} has been introduced as a double-sided RIS, which additionally enables the re-transmission of incident signals to the rear zones of the RIS, resulting in a full-space 360-degree coverage for all users. In addition to introducing another degree of freedom (DoF) for double-sided RIS interaction, a STAR-RIS promotes the EE of conventional one-sided RIS-based systems~\cite{STAR-Surv}. To date, a large number of articles have investigated STAR-RIS supported communications from the point of view of radio resource management. In~\cite{STAR-Impair}, active STAR-RIS was introduced by adopting an efficient gain factor and further gains over the previous unidirectional RIS-assisted systems were numerically demonstrated. 
The energy-efficient design of a STAR-RIS-aided system with non-orthogonal multiple access (NOMA) was studied in~\cite{Ref1_STAR-RIS}, where the authors invoked classical convex optimisation techniques to jointly control the transmit beamforming at the BS and the phase shift at the STAR-RIS. A novel large-scale fading decoding scheme was proposed in~\cite{Ref2_STAR-RIS} to improve the spectral efficiency of an active STAR-RIS-supported cell-free massive multiple-input multiple-output (MIMO) system. Numerical results demonstrated a significant improvement for the proposed system over the conventional active RIS-aided cell-free massive MIMO system.
\textcolor{black}{The authors of \cite{learning1} and \cite{learning2} investigated the potential of deep reinforcement learning (DRL) for optimising the performance of STAR-RIS-aided systems. Specifically, in \cite{learning1}, an online joint active and passive beamforming framework was proposed to maximise the long-term EE of a multi-cell STAR-RIS-enabled system using a parallel DRL technique. In \cite{learning2}, the maximisation of the overall EE of a NOMA-enabled STAR-RIS-based network was investigated using a deep deterministic policy gradient (DDPG)-based DRL algorithm.} 
\textit{Concerning the massive scalability and also energy limitation of IoT devices, an active STAR-RIS-aided SWIPT turns out to be a potential technology for future IoT networks, which has never been studied up to now.}

\subsection{Research Challenges and Contributions}
\par Despite the many advantages mentioned above, the optimisation of a SWIPT-enabled active STAR-RIS-assisted system raises several challenges. In particular, the optimisation of the transmit beamforming at the BS and the PS ratio of the IoT devices, as well as the phase shift, gain and element selection at the STAR-RIS are of paramount importance in such a system. This system is therefore complex due to the coupling of various optimisation variables. So far, two main categories of resource allocation schemes have been investigated for optimising a STAR-RIS based system, namely (i) the classical convex optimisation based ones, e.g. \cite{optimization1,optimization2}, and (ii) the recent learning-driven ones, e.g. \cite{learning1,learning2}.
The former category applies mathematical transformations to reformulate the non-convex resource management optimisation problems in convex form and address them via existing off-the-shelf solvers, e.g., CVX. In contrast, DRL, as the prominent learning framework belonging to the latter category, provides a real-time and joint optimisation framework, provided that the size of its action space is affordable. 
For resource allocation optimisation problems with numerous highly coupled and scalable variables, DRL-based methods, as standalone optimisers, encounter severe performance degradation due to the exponential order of their action space size. Meanwhile, the classical convex optimisation-based methods face high computational complexity due to a large number of mathematical computations~\cite{AARIS_paper}.

In this paper, we propose a resource allocation framework by exploiting the potentials of the two aforementioned solution strategies to evaluate the performance of a SWIPT-enabled active STAR-RIS-assisted communication system. Furthermore, we propose an element selection mechanism to optimise the on/off status of active STAR-RIS elements and manage their energy consumption. Specifically, the key contributions of this study can be categorised as follows:
\begin{itemize}
	\item An active STAR-RIS-based system is studied in a downlink transmission, where a single-antenna user employs non-linear SWIPT technology for superimposing energy and information in the power domain of the transmitted signal. Our work differs from \cite{STAR-Swipt} by adopting active beamforming at the STAR-RIS and also differs from \cite{STAR-Impair} by employing SWIPT technology. In addition, an energy-efficient element selection mechanism is designed in this paper to mitigate the significant energy consumption of the active STAR-RIS system, in contrast to~\cite{STAR, STAR-Surv} and \cite{Ref1_STAR-RIS, Ref2_STAR-RIS, learning1, learning2, optimization1,optimization2}.
	Under the assumption of weak LoS links between the BS and the users, an active STAR-RIS is deployed, which uses active beamforming to redirect (transmit or reflect) incident signals and simultaneously serves the users with energy and information. 
	\item A resource allocation optimisation problem is formulated to maximise the overall system EE by jointly optimising the gain, element selection and phase shift at the active STAR-RIS, as well as the PS ratio of the users and the transmit beamforming at the BS. This problem is subject to the BS transmit power budget, as well as a minimum signal-to-interference-plus-noise ratio (SINR) and EH requirements for all users. The problem is complicated by the fractional form of the EE and the inclusion of highly coupled variables. Consequently, an alternating optimisation framework is proposed by iteratively invoking tools from convex optimisation and DRL.
	\textcolor{black}{ In particular, a solution based on convex optimisation is proposed to jointly optimise the transmit beamforming at the BS as well as the PS ratio of all users. First, a semidefinite relaxation (SDR) method drops the BS beamforming rank constraints. Then, a difference-of-concave-function (DC) method followed by a first-order Taylor expansion is applied for more tractability. Due to the fractional form of the EE, we decompose it into an equivalent linear function using the Dinkelbach method, which transforms the problem into a standard convex semidefinite programming (SDP) problem. Finally, a standard solver jointly optimises the PS ratio of the users and the transmit beamforming at the BS. We then develop a novel DRL-based algorithm called modified deep deterministic policy gradient and soft actor-critic (MDS). This algorithm jointly uses a modified DDPG~\cite{DDPG} to optimise the element selection at the active STAR-RIS, as well as a soft actor-critic (SAC)~\cite{Soft} to optimise its gain factor and phase shift. We further enhance the performance of the proposed MDS algorithm by integrating it with a model-agnostic meta-learning scheme~\cite{MAML}. Compared to \cite{learning1,learning2}, our enhanced meta-MDS (MMDS) algorithm performs well in highly dynamic scenarios with user mobility by enhancing the generalisation ability of the MDS algorithm and showing fast adaptability to the upcoming network dynamics.}
	\item Simulation results show that our proposed active STAR-RIS-based SWIPT scheme achieves a significant system EE gain of around $57\%$, compared to its passive counterpart. In addition, the proposed MMDS scheme, by integrating model-agnostic meta-learning, achieves a system EE gain of around $32\%$, compared to the originally introduced MDS scheme. This gain is achieved due to the better adaptability and generalisation capability of the proposed MMDS scheme.
\end{itemize}
The remainder of this paper is organized as follows. In Section \ref{system model}, we introduce the system model. Then, the EE maximization problem is formally stated, along with a convexity analysis and a big picture of the solution strategy in Section \ref{problem statement}. The classical optimisation method for jointly optimising the transmit beamforming at the BS, as well as the PS ratio of all users  is described in Section \ref{SectionIV}. The proposed MDS algorithm is explained in Section \ref{MDS}. In Section \ref{MMDS}, the MMDS algorithm is proposed. Finally, simulation results and conclusions are given in Sections \ref{simulation} and \ref{Conclusion}, respectively.
\par \textit{Notations}: In this paper, the following notations are used. The capital boldface letters represent matrices, while the small bold and small normal letters denote vectors and scalars, respectively. An identity matrix is denoted by $\mathbf{I}$. Additionally, $\mathbb{C}^{x\times y}$ signifies an $x\times y$ complex-valued matrix, and the superscript $H$ stands for the complex conjugate transpose of a matrix. 
Notations $\mathbb{E}\left[\cdot \right]  $ and
$\mathrm{diag}(\cdot)$ are used for the statistical expectation and diagonalization operator, respectively.
$||\mathbf{x}||$ corresponds to the 2-norm of a vector $\mathbf{x}$, and $\lVert \mathbf{A} \rVert_{F}$ designates the Frobenius norm of the matrix $\mathbf{A}$ with an
arbitrary size.
\vspace*{-1em}
\section{System Model}\label{system model}
\vspace*{-0.5em}
\begin{figure} 	
	\centering
	\includegraphics[scale=0.35]{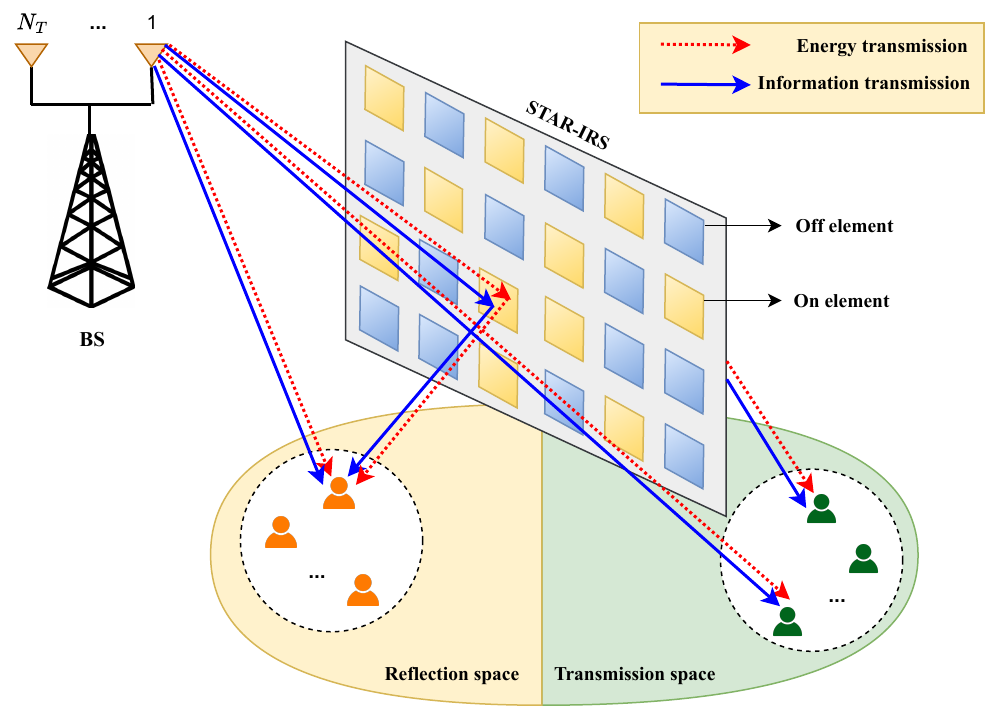}
	\caption{An active STAR-RIS-aided system.}
	\label{system_model} 
\end{figure}

As depicted in Fig. \ref{system_model}, we consider the downlink transmission of a multiple-input-single-output (MISO) system, where a BS with $N_{\textrm{t}}$ antennas serves single-antenna users for both information and energy at the same time via SWIPT technology. In order to extend the coverage of the network and facilitate the communication, an active STAR-RIS is deployed between the BS and users, which strengthens the incident signals via an amplification mechanism and provides the system with virtual LoS links. It is assumed that $M$ reconfigurable elements are deployed on the active STAR-RIS, capable of modifying the phase of the BS signals independently. Accordingly, based on their relative position to the active STAR-RIS, users are categorized into two sets: a set $\mathcal{U}_{\textrm{r}} = \lbrace 1, 2, \dots, U_{\textrm{r}} \rbrace$ of $U_{\textrm{r}}$ users distributed in front of the active STAR-RIS, and another set $\mathcal{U}_{\textrm{t}} = \lbrace 1, 2, \dots, U_{\textrm{t}} \rbrace$ of $U_{\textrm{t}}$ users located behind it. In such a scenario, the set of users in $\mathcal{U}_{\textrm{t}}$ receive the retransmitted signal by the active STAR-RIS, whereas the set of users in $\mathcal{U}_{\textrm{r}}$ receive its reflected signal. The signals communicate through quasi-static flat-fading channels. Similar to \cite{Hosein5}, the perfect channel state information (CSI) is assumed to be available at the transmitter (i.e., the BS), by leveraging efficient channel estimation mechanisms \cite{channel_estimation1}, \cite{channel_estimation2}, \cite{channel_estimation3}. Accordingly, the transmit signal from the BS can be expressed as
\begin{equation}\label{eq1}
\mathbf{x} =  \sum_{i \in \mathcal{U}_{\textrm{r}}} \mathbf{w}_{\textrm{r},i} s_{i} + \sum_{j \in \mathcal{U}_{\textrm{t}}} \mathbf{w}_{\textrm{t}, j} s_{j}, 
\end{equation}
in which $\mathbf{w}_{\textrm{r}, i} \in \mathbb{C}^{N_{\textrm{t}} \times 1}$ and $\mathbf{w}_{\textrm{t}, j} \in \mathbb{C}^{N_{\textrm{t}} \times 1}$ denote the beamforming vectors for users $i$ and $j$, located in the reflection and re-transmission zones of the active STAR-RIS, respectively. 
Similarly, the information symbols for users $i$ and $j$ are indicated by $s_{i}$ and $s_{j}$, where $s_{i} \thicksim \mathcal{CN}(0,1), \forall i \in \mathcal{U_{\textrm{t}}}$ and $s_{j} \thicksim \mathcal{CN}(0,1), \forall i \in \mathcal{U_{\textrm{r}}}$.
  
\vspace*{-0.2em}
We express the channel matrix between the BS and  the active STAR-RIS as $\mathbf{G}\in%
\mathbb{C}
^{M\times N_{\textrm{t}}}$. The direct channel vectors from the BS to users $i\in \mathcal{U_{\textrm{r}}}$ and $j\in \mathcal{U_{\textrm{t}}}$ are denoted by ${\mathbf{h}^{H}_{\textrm{r}, i, \textrm{d}}}\in
\mathbb{C}
^{1 \times N_{\textrm{t}}}$ and  ${\mathbf{h}^{H}_{\textrm{t}, j, \textrm{d}}} \in
\mathbb{C}
^{1 \times N_{\textrm{t}}}$, respectively. Similarly, we let ${\mathbf{h}^{H}_{\textrm{r},i}}\in
\mathbb{C}
^{1 \times M}$ and ${\mathbf{h}^{H}_{\textrm{t},j}}\in
\mathbb{C}
^{1 \times M}$, respectively, indicate the reflecting and re-transmitting channel vectors from the active STAR-RIS to users $i\in \mathcal{U_{\textrm{r}}}$ and $j\in \mathcal{U_{\textrm{t}}}$. 
The active STAR-RIS is capable of adjusting its amplification factor matrix $\mathbf{A}=\mathrm{diag}\left(  \mathbf{a}\right) \in \mathbb{R}_{+}^{M \times M} $, with $\mathbf{a} = [a_{1},a_{2} ,...,a_{M}]$, as well as its phase shift matrix $\mathbf{\Theta}_{\textrm{r}}=$$\mathrm{diag}$$\left( \theta_{\textrm{r}, 1}, \dots, \theta_{\textrm{r}, M} \right) \in \mathbb{C}^{M \times M}$ and  $\mathbf{\Theta}_{\textrm{t}}=$$\mathrm{diag}$$\left( \theta_{\textrm{t},1}, \dots, \theta_{\textrm{t}, M} \right) \in \mathbb{C}^{M \times M} $, for the reflection and re-transmission, respectively. In particular, for a typical element $m \in \mathcal{M} \triangleq \lbrace 1, \dots, M\rbrace$ at the active STAR-RIS, $\theta_{\textrm{t}, m}=\sqrt{\beta_{\textrm{t}, m}}e^{\textrm{j} \varphi_{\textrm{t}, m}}$ stands for the re-transmission coefficient with $\sqrt{\beta_{\textrm{t}, m}} \in [0, 1]$ and $\varphi_{\textrm{t}, m}$ $\in$ $[0,2\pi)$ being its amplitude and phase shift, respectively. Likewise, the reflection coefficient of a typical element $m$ at the active STAR-RIS is modeled as  $\theta_{\textrm{r}, m}=\sqrt{\beta_{\textrm{r}, m}}e^{\textrm{j}\varphi_{\textrm{r}, m}}$ in which $\sqrt{\beta_{\textrm{r}, m}} \in [0, 1]$ and $\varphi_{\textrm{r}, m}$ $\in$ $[0,2\pi)$, respectively,
represent its amplitude and phase shift. 
Taking the energy conservation law into account, the total energy of the re-transmitted and reflected signals from the same element $m$ at the active STAR-RIS is constrained to $\beta_{\textrm{t}, m} + \beta_{\textrm{r}, m} = 1$.
Hence, the received signal at user $i\in \mathcal{U}_{\textrm{r}}$ in reflection side of the STAR-RIS can be expressed as
\begin{subequations}\label{eq2}
	\begin{align}
	y_{\textrm{r}, i} & ={\mathbf{h}^{H}_{\textrm{r}, i, \textrm{d}}} \mathbf{x} + {\mathbf{h}^{H}_{\textrm{r}, i}} \mathbf{A \Theta}_{\textrm{r}} \big(\mathbf{G x} + \mathbf{z}_{\textrm{r}}\big) + n_{\textrm{r}, i} \label{eq2_1}\\
	& = \big({\mathbf{h}^{H}_{\textrm{r}, i, \textrm{d}}} + {\mathbf{h}^{H}_{\textrm{r}, i}} \mathbf{A\Theta}_{\textrm{r}} \mathbf{G}\big) \mathbf{x} + {\mathbf{h}^{H}_{\textrm{r}, i}}\mathbf{A\Theta}_{\textrm{r}} \mathbf{z}_{\textrm{r}} + n_{\textrm{r}, i} \label{eq2_2}, \quad i \in \mathcal{U}_{\textrm{r}},
	\end{align}
\end{subequations}
where $\mathbf{z}_{\textrm{r}} \in \mathcal{C}^{M \times 1} \thicksim \mathcal{CN}(\mathbf{0}_{M},{\sigma^{2}_{\textrm{r}, z}} \mathbf{I}_{M})$ and $n_{\textrm{r}, i} \thicksim \mathcal{CN}(0,{\sigma^{2}_{\textrm{r}, i}})$ stand for the noise, caused by active STAR-RIS and additive white Gaussian noise (AWGN) at user $i$, with corresponding variances ${\sigma^{2}_{\textrm{r}, z}}$ and ${\sigma^{2}_{\textrm{r}, i}}$, respectively. 
\par Concerning the energy consumption of active beamforming at the STAR-RIS, an element selection mechanism is crucial to determine the elements' on/off status, depending on situation. This mechanism enhances the EE of the system significantly~\cite{ZhuHan}. To this aim, the binary element selection matrix $\mathbf{F}$ is introduced as $\mathbf{F} = \text{diag}(\mathbf{f}) \in \lbrace 0, 1 \rbrace^{M \times M}$ where $\mathbf{f} = [f_{1}, \dots, f_{M}]^T \in \lbrace 0, 1 \rbrace^{M \times 1}$, implying that the $m$-th element of the active STAR-RIS is on for $f_{i}=1$, and off, otherwise. Assume that $N$ elements ($1 \le N \le M$) of the active STAR-RIS are active, while the rest are inactive due to energy considerations. Then, by taking into account the element selection matrix $\mathbf{F}$, \eqref{eq2} can be rewritten as
\begin{equation}\label{received_signal_element_selection}
\begin{split}
y_{\textrm{r}, i}  =& \left({\mathbf{h}^{H}_{\textrm{r}, i, \textrm{d}}} + {\mathbf{h}^{H}_{\textrm{r}, i}} \mathbf{A} \mathbf{F \Theta_{\textrm{r}}} \mathbf{G}\right) \mathbf{x} +  {\mathbf{h}^{H}_{\textrm{r},i}} \mathbf{A} \mathbf{F \Theta_{\textrm{r}}}  \mathbf{z}_{\textrm{r}} + n_{\textrm{r}, i} \\
=& ~\bar{\mathbf{h}}^{H}_{\textrm{r}, i} \mathbf{x} +  {\mathbf{h}^{H}_{\textrm{r}, i}} \mathbf{A} \mathbf{F \Theta_{\textrm{r}}}  \mathbf{z}_{\textrm{r}} + n_{\textrm{r}, i} , \quad i \in \mathcal{U}_{\textrm{r}},
\end{split}
\end{equation}
in which  $\bar{\mathbf{h}}^{H}_{\textrm{r}, i} = \left({\mathbf{h}^{H}_{\textrm{r}, i, \textrm{d}}} + {\mathbf{h}^{H}_{\textrm{r}, i}} \mathbf{A} \mathbf{F \Theta_{\textrm{r}}} \mathbf{G}\right)$ signifies the equivalent end-to-end channel from the BS to the user $i$. As discussed earlier, the BS as the transmitter employs SWIPT technology for accomodating the energy and information within a unified signal using the power domain. At the receiver side, the user $i$ is equipped with a PS mechanism for ID and EH at the same time, respectively expressed as
\vspace*{-1em}

\small
\begin{equation} \label{rho1}
y_{\textrm{r}, i}^{\text{ID}} = \sqrt{\rho_{\textrm{r}, i}} \left( \bar{\mathbf{h}}^{H}_{\textrm{r}, i} \mathbf{x} + {\mathbf{h}^{H}_{\textrm{r}, i}} \mathbf{A} \mathbf{F \Theta_{\textrm{r}}}  \mathbf{z}_{\textrm{r}} + n_{\textrm{r}, i}   \right) + \eta_{\textrm{r}, i}, \ \ \forall i \in \mathcal{U}_{\textrm{r}},
\end{equation}
\normalsize
\vspace*{-1em}
and
\vspace*{-1em}

\small
\begin{equation} \label{rho2}
y_{\textrm{r}, i}^{\text{EH}} = \sqrt{1 - \rho_{\textrm{r}, i}} \left( \bar{\mathbf{h}}^{H}_{\textrm{r}, i} \mathbf{x} + {\mathbf{h}^{H}_{\textrm{r}, i}} \mathbf{A} \mathbf{F \Theta_{\textrm{r}}} \mathbf{z}_{\textrm{r}} + n_{\textrm{r}, i}   \right) , \ \ \forall i \in \mathcal{U}_{\textrm{r}} ,
\end{equation}
\normalsize
such that ${\eta_{\textrm{r}, i}} \sim \mathcal{C}\mathcal{N}\left( {0,{\delta^{2}_{\textrm{r}, i}}} \right)$ denotes the additional noise as a result of
the signal processing at the ID receiver. Moreover, $0 \le {{{\rho_{\textrm{r}, i}}}} \le 1$ in (\ref{rho1}) determines the portion of the received signal, corresponding to ID, whereas the remaining portion, i.e., $1 - {{{\rho_{\textrm{r}, i}}}}$ in (\ref{rho2}) is dedicated to EH. Therefore, the received signal-to-interference-plus-noise-ratio (SINR) and the harvested energy of the user $i \in \mathcal{U}_{\textrm{r}}$ can be defined as
\vspace*{-1.5em}

\small
\begin{equation}\label{eq_3}
\gamma_{\textrm{r}, i}\!\!=\!\!\frac{\rho_{\textrm{r}, i} \lvert\bar{\mathbf{h}}^{H}_{\textrm{r},i} \mathbf{w}_{\textrm{r}, i}\rvert^{2}}{%
	{\rho_{\textrm{r}, i} \textstyle\sum\nolimits_{k \in \mathcal{U}_{\textrm{r}},k\neq i}}
	\lvert\bar{\mathbf{h}}^{H}_{\textrm{r},i} \mathbf{w}_{\textrm{r}, k}\rvert^{2} \!\!+\!\! \rho_{\textrm{r}, i} {\sigma^2_{\textrm{r}, z}}\lVert {\mathbf{h}^{H}_{\textrm{r}, i}} \mathbf{A} \mathbf{F \Theta_{\textrm{r}}}  \rVert^{2} \!\!+\!\! \rho_{\textrm{r}, i} {\sigma^2_{\textrm{r}, i}} \!\!+\!\! {\delta^2_{\textrm{r}, i}}},
\end{equation}
\normalsize
for $\forall i \in \mathcal{U}_{\textrm{r}}$ and 
\vspace*{-1em}

\small
\begin{equation}\label{eq_4}
P_{\textrm{r}, i} =  \left(  1-\rho_{\textrm{r}, i} \right) \left( \sum\nolimits_{k \in \mathcal{U}_{\textrm{r}}}
\lvert\bar{\mathbf{h}}^{H}_{\textrm{r}, i} \mathbf{w}_{\textrm{r}, k} \rvert^{2} \!\!+\!\! {\sigma^2_{\textrm{r}, z}}\lVert {\mathbf{h}^{H}_{\textrm{r}, i}} \mathbf{A} \mathbf{F \Theta_{\textrm{r}}}  \rVert^{2}  \right), \ \ \forall i \in \mathcal{U}_{\textrm{r}}.
\end{equation}
\vspace*{-0.3em}
\normalsize
Similar to users in $\mathcal{U}_{\textrm{r}}$, the received signal at user $j\in \mathcal{U}_{\textrm{t}}$ in re-transmission side of the active STAR-RIS, is given by
\begin{align}\label{ten}
y_{\textrm{t}, j} & = \left({\mathbf{h}^{H}_{\textrm{t},j, \textrm{d}}} + {\mathbf{h}^{H}_{\textrm{t}, j}} \mathbf{A} \mathbf{F \Theta_{\textrm{t}}} \mathbf{G}\right) \mathbf{x} +  {\mathbf{h}^{H}_{\textrm{t}, j}} \mathbf{A} \mathbf{F \Theta_{\textrm{t}}}  \mathbf{z}_{\textrm{t}} + n_{\textrm{t}, j} \  \\
& = ~\bar{\mathbf{h}}^{H}_{\textrm{t}, j}  \mathbf{x} +  {\mathbf{h}^{H}_{\textrm{t}, j}} \mathbf{A} \mathbf{F \Theta_{\textrm{t}}}  \mathbf{z}_{\textrm{t}} + n_{\textrm{t}, j} , \ \  j \in \mathcal{U}_{\textrm{t}}
\end{align}
where $\mathbf{z}_{\textrm{t}} \thicksim \mathcal{CN}(\mathbf{0}_{M},{\sigma^{2}_{\textrm{t}, z}} \mathbf{I}_{M})$ stands for the noise caused by active STAR-RIS with corresponding variance $\sigma^{2}_{\textrm{t}, z}$, while $n_{\textrm{t}, j} \thicksim \mathcal{CN}(0,{\sigma^{2}_{\textrm{t}, j}})$ denotes the AWGN with corresponding variance ${\sigma^{2}_{\textrm{t}, j}}$ for user $j$. Moreover, $\bar{\mathbf{h}}^{H}_{\textrm{t}, j}  = \left({\mathbf{h}^{H}_{\textrm{t}, j, \textrm{d}}} + {\mathbf{h}^{H}_{\textrm{t}, j}} \mathbf{A} \mathbf{F \Theta_{\textrm{t}}} \mathbf{G}\right)$ represents the equivalent end-to-end channel from the BS to user $j$. With the aid of a PS mechanism, this user is capable of simultaneously benefiting from ID and EH, respectively given by

\small
\begin{equation} \label{rho3}
y_{\textrm{t}, j}^{\text{ID}} = \sqrt{\rho_{\textrm{t}, j}} \left( \bar{\mathbf{h}}^{H}_{\textrm{t}, j} \mathbf{x} + {\mathbf{h}^{H}_{\textrm{t}, j}} \mathbf{A} \mathbf{F \Theta_{\textrm{t}}}  \mathbf{z}_{\textrm{t}} + n_{\textrm{t}, j}   \right) + \eta_{\textrm{t}, j}, \ \  \forall j \in \mathcal{U}_{\textrm{t}} ,
\end{equation}
\normalsize
and
\vspace*{-1em}

\small
\begin{equation} \label{rho4}
y_{\textrm{t}, j}^{\text{EH}} = \sqrt{1 - \rho_{\textrm{t}, j}} \left( \bar{\mathbf{h}}^{H}_{\textrm{t}, j} \mathbf{x} + {\mathbf{h}^{H}_{\textrm{t}, j}} \mathbf{A} \mathbf{F \Theta_{\textrm{t}}} \mathbf{z}_{\textrm{t}} + n_{\textrm{t}, j}   \right) , \ \ \forall j \in \mathcal{U}_{\textrm{t}} ,
\end{equation}
\normalsize
in which ${\eta_{\textrm{t}, j}} \sim \mathcal{C}\mathcal{N}\left( {0,{\delta^2_{\textrm{t}, j}}} \right)$. The portion of the received signal for ID in \eqref{rho3} is denoted by $0 \le {{{\rho_{\textrm{t}, j}}}} \le 1$. In contrast, the remaining portion of the signal is utilized for EH in \eqref{rho4}, which is expressed as $1 - {{{\rho_{\textrm{t}, j}}}}$. Accordingly, the received SINR and the harvested energy of the user $j \in \mathcal{U}_{\textrm{t}}$ are determined as
\vspace*{-1.5em}

\small
\begin{equation}\label{eq_33}
\gamma_{\textrm{t}, j}\!\!=\!\!\dfrac{\rho_{\textrm{t}, j} \lvert\bar{\mathbf{h}}^{H}_{\textrm{t}, j}\mathbf{w}_{\textrm{t}, j}\rvert^{2}}{%
	{\rho_{\textrm{t}, j} \textstyle\sum\nolimits_{k \in \mathcal{U}_{\textrm{t}},k\neq j}}
	\lvert\bar{\mathbf{h}}^{H}_{\textrm{t}, j} \mathbf{w}_{\textrm{t}, k}\rvert^{2} \!\!+\!\! \rho_{\textrm{t}, j} {\sigma^{2}_{\textrm{t}, z}}\lVert {\mathbf{h}^{H}_{\textrm{t}, j}} \mathbf{A} \mathbf{F \Theta_{\textrm{t}}}  \rVert^{2} \!\!+\!\! \rho_{\textrm{t}, j} {\sigma^2_{\textrm{t}, j}} \!\!+\!\! {\delta^{2}_{\textrm{t}, j}}},
\end{equation}
\normalsize
for $\forall j \in \mathcal{U}_{\textrm{t}}$ and 

\small
\begin{equation}\label{eq_44}
P_{\textrm{t}, j} =  \left(  1-\rho_{\textrm{t}, j} \right) \left( \sum\nolimits_{k \in \mathcal{U}_{\textrm{t}}}
\lvert\bar{\mathbf{h}}^{H}_{\textrm{t}, j}\mathbf{w}_{\textrm{t}, k}\rvert^{2} \!\!+\!\! {\sigma^{2}_{\textrm{t}, z}}\lVert {\mathbf{h}^{H}_{\textrm{t}, j}} \mathbf{A} \mathbf{F \Theta_{\textrm{t}}}  \rVert^{2}  \right), \ \ \forall j \in \mathcal{U}_{\textrm{t}}.
\end{equation}
\normalsize
respectively. Finally, based on Shannon-Hartley theorem~\cite{capacity}, the total achievable information rate of all users is given by
\begin{equation}\label{rate}
R(\mathcal{H}_{1}) \triangleq \sum_{i \in \mathcal{U}_{\textrm{r}}} \log_{2}(1+\gamma_{\textrm{r}, i})  + \sum_{j \in \mathcal{U}_{\textrm{t}}} \log_{2}(1+\gamma_{\textrm{t}, j}).
\end{equation}
where $\mathcal{H}_{1} = \Big(\lbrace \mathbf{w}_{\textrm{r}, i} \rbrace_{i \in \mathcal{U}_{\textrm{r}}}, \lbrace \mathbf{w}_{\textrm{t}, j} \rbrace_{j \in \mathcal{U}_{\textrm{t}}},\mathbf{A}, \mathbf{\Theta_{\textrm{r}}}, \mathbf{\Theta_{\textrm{t}}}, \mathbf{F}, \lbrace \rho_{\textrm{r}, i}\rbrace_{i \in \mathcal{U}_{\textrm{r}}},$
$ \lbrace\rho_{\textrm{t}, j} \rbrace_{j \in \mathcal{U}_{\textrm{t}}}\Big)$. However, a linear EH model does not capture the effect of the non-linear behavior of circuits and therefore is not practical. Therefore, we adopt a non-linear EH model for a typical user $i\in \mathcal{U}_{\textrm{r}}$, which is $E_{\textrm{r}, i}^{\text{NL}} = \frac{{\Psi _{\textrm{r}, i}^{\text{NL}} - {M_{\textrm{r}, i}}{\Omega _{\textrm{r}, i}}}}{{1 - {\Omega _{\textrm{r}, i}}}}$  ~\cite{non-linear-EH},~\cite{es9}
with
\small
\begin{equation}\label{PSI_r}
\Psi _{\textrm{r}, i}^{\text{NL}} = \frac{{{M_{\textrm{r}, i}}}}{{1 + \exp ( - {a_{\textrm{r}, i}}({P_{\textrm{r}, i}} - {b_{\textrm{r}, i}}))}}, \quad {\Omega _{\textrm{r}, i}} = \frac{1}{{1 + \exp ({a_{\textrm{r}, i}}{b_{\textrm{r}, i}})}},
\end{equation}
\normalsize
where $\Psi _{\textrm{r}, i}^{\text{NL}}$ represents the traditional logistic function and ${\Omega _{\textrm{r}, i}}$ is a constant to guarantee a zero input/output response.
Furthermore, ${a_{\textrm{r}, i}}$ and ${b_{\textrm{r}, i}}$ correspond to constants related to the circuit characteristics, while ${{M_{\textrm{r},i}}}$ specifies a constant indicating the maximum harvested energy at the active STAR-RIS.
Likewise, for a typical user $j\in \mathcal{U}_{\textrm{t}}$, the non-linear EH model is given by $E_{\textrm{t}, j}^{\text{NL}} = \frac{{\Psi _{\textrm{t}, j}^{\text{NL}} - {M_{\textrm{t}, j}}{\Omega _{\textrm{t}, j}}}}{{1 - {\Omega _{\textrm{t}, j}}}}$ with
\small
\begin{equation}\label{PSI_t}
\Psi _{\textrm{t}, j}^{\text{NL}} = \frac{{{M_{\textrm{t}, j}}}}{{1 + \exp ( - {a_{\textrm{t}, j}}({P_{\textrm{t}, j}} - {b_{\textrm{t}, j}}))}}, \quad {\Omega _{\textrm{t}, j}} = \frac{1}{{1 + \exp ({a_{\textrm{t}, j}}{b_{\textrm{t}, j}})}}.
\end{equation}
\normalsize
Other corresponding retransmission constants ${b_{\textrm{t},j}}$, ${b_{\textrm{t}, j}}$ and ${{M_{\textrm{t}, j}}}$ are defined following their reflective counterparts. On the basis of the abovementioned non-linear EH, resources have to be allocated, such that the harvested energy of users $i\in \mathcal{U}_{\textrm{r}}$ and $j\in \mathcal{U}_{\textrm{t}}$, satisfy minimum EH thresholds ${E_{\textrm{r}, i}^{\textrm{min}}}$ and ${E_{\textrm{t}, j}^{\textrm{min}}}$, respectively. These constraints are formally defined as $E_{\textrm{r}, i}^{\text{NL}}\ge{E_{\textrm{r}, i}^{\textrm{min}}}$ and $E_{\textrm{t}, j}^{\text{NL}}\ge{E_{\textrm{t}, j}^{\textrm{min}}}$.
In wireless networks, a well-known definition of EE is the ratio of total achievable data rate (obtained from \eqref{rate}) and total power consumption of the system which is modeled as
\begin{equation}\label{total power}
P(\mathcal{H}_{2}) \triangleq \sum_{i \in \mathcal{U}_{\textrm{r}}} \lVert \mathbf{w}_{\textrm{r}, i} \rVert^{2} +  \sum_{j \in \mathcal{U}_{\textrm{t}}} \lVert \mathbf{w}_{\textrm{t}, j} \rVert^{2} + P_{\mathrm{Cir}} + P_{\mathrm{ASR}},
\end{equation}

\vspace*{-0.3em}
\noindent where $\mathcal{H}_{2} = \Big(\lbrace \mathbf{w}_{\textrm{r}, i} \rbrace_{i \in \mathcal{U}_{\textrm{r}}}, \lbrace \mathbf{w}_{\textrm{t}, j} \rbrace_{j \in \mathcal{U}_{\textrm{t}}} ,\mathbf{A}, \mathbf{F}, \mathbf{\Theta_{\textrm{r}}}, \mathbf{\Theta_{\textrm{t}}}\Big)$. Also, $P_{\mathrm{Cir}} = P_{\mathrm{BS}}^{\mathrm{Cir}} + \sum_{i \in \mathcal{U}_{\textrm{r}}} p_{i}^{\mathrm{Cir}} + \sum_{j \in \mathcal{U}_{\textrm{t}}} p_{j}^{\mathrm{Cir}}$ specifies the constant circuit power dissipation, including the power consumption of the BS and users, denoted by $P_{\mathrm{BS}}^{\mathrm{Cir}}$ and $p_{i}^{\mathrm{Cir}}$$(p_{j}^{\mathrm{Cir}})$, respectively. Moreover, in (\ref{total power}), the total power consumption for the active STAR-RIS is stated as $P_{\mathrm{ASR}} = N(P_{\mathrm{c}} + P_{\mathrm{DC}}) + \zeta p_{\mathrm{out}}$, in which $P_{\mathrm{c}}$ represents the power consumption of the switch and control circuit~\textcolor{black}{at each element}, $P_{\mathrm{DC}}$ corresponds to the biasing power consumption~\textcolor{black}{at each element} and $\zeta \triangleq \nu^{-1}$, with $\nu$ being the amplifier efficiency~\cite{ACTIVE-STAR-RIS}. In addition, $p_{\mathrm{out}}$ stands for the output power of the active STAR-RIS, expressed as $p_{\mathrm{out}} = \sum_{i \in \mathcal{U}_{\textrm{r}}} \lVert \mathbf{A F} \mathbf{\Theta_{\textrm{r}}} \mathbf{G} \mathbf{w}_{\textrm{r}, i}   \rVert^{2} + \sum_{j \in \mathcal{U}_{\textrm{t}}} \lVert \mathbf{A F} \mathbf{\Theta_{\textrm{t}}} \mathbf{G} \mathbf{w}_{\textrm{t}, j}   \rVert^{2}  + {\sigma^{2}_{\textrm{r}, z}} \lVert \mathbf{A F} \mathbf{\Theta_{\textrm{r}}} \rVert_{F}^{2} + {\sigma^{2}_{\textrm{t}, z}} \lVert \mathbf{A F} \mathbf{\Theta_{\textrm{t}}} \rVert_{F}^{2}$ ~\cite{ACTIVE-STAR-RIS}.

\vspace*{-1em}
\section{Optimisation Problem}\label{problem statement}
In this section, aiming at maximizing the overall system EE, we formulate a problem, which jointly optimises the beamforming vector and PS ratio of all users, as well as the amplification factor, phase shift and element selection matrix at the active STAR-RIS. 
\subsection{Problem Definition}
Based on the system model introduced in the previous section, we mathematically formulate the network-wide optimisation problem in the system model we discussed earlier as follows
\begin{subequations}\label{eq5}
	\begin{align}
	\mathcal{P}: \ \ &  \max_{\mathcal{H}_{1}}\ \ \quad
	\dfrac{R(\mathcal{H}_{1})}{P(\mathcal{H}_{2})} \\ 
	\mathrm{s.t.} \ \  &   \text{C}_{1}: \gamma_{\textrm{r}, i} \geq \gamma^{\min}_{\textrm{r}, i}, ~~~\forall i\in \mathcal{U}_{\textrm{r}}, \quad \\
	&   \text{C}_{2}:  \gamma_{\textrm{t}, j} \geq \gamma^{\min}_{\textrm{t}, j}, ~~~\forall j\in \mathcal{U}_{\textrm{t}}, \quad \\
	&   \text{C}_{3}: E_{\textrm{r}, i}^{\text{NL}} \ge {E_{\textrm{r}, i}^{\min}} ,  ~~~\forall i\in \mathcal{U}_{\textrm{r}} , \quad \\
	&   \text{C}_{4}: E_{\textrm{t}, j}^{\text{NL}} \ge {E_{\textrm{t}, j}^{\min}} ,  ~~~\forall j\in \mathcal{U}_{\textrm{t}} , \quad \\
	&    \text{C}_{5}: \sum_{i \in \mathcal{U}_{\textrm{r}}} \lVert \mathbf{w}_{\textrm{r}, i} \rVert^{2} + \sum_{j \in \mathcal{U}_{\textrm{t}}} \lVert \mathbf{w}_{\textrm{t}, j} \rVert^{2} \leq P_{\mathrm{BS}}^{\max}, \quad \\
	&   \text{C}_{6}: p_{\mathrm{out}} \leq {P_{I}^{\max}}, \label{eq5_5} \quad \\
	&  \text{C}_{7}:  0 \le {\rho_{\textrm{r}, i}} \le 1, \quad \quad \forall i \in \mathcal{U}_{\textrm{r}}, \quad  \\
	& \text{C}_{8}: 0 \le {\rho_{\textrm{t}, j}} \le 1, \quad \quad \forall j \in \mathcal{U}_{\textrm{t}}, \quad \\
	&  \text{C}_{9}: 0 \leq a_{m}  \leq a_{\max}, \quad \ \ \ \forall m \in \mathcal{M}, \quad \\
	&  \text{C}_{10}: f_{m} \in \lbrace 0,1 \rbrace, \quad \ \ \  \forall m \in \mathcal{M}, \quad \\
	& \textcolor{black}{\text{C}_{11}: \lVert \mathbf{F} \rVert_{F} \leq N,} \quad  \\
	& \textcolor{black}{\text{C}_{12}: \beta_{\textrm{r}, m}, \beta_{\textrm{t}, m} \in [0, 1], \ \ \beta_{\textrm{r}, m} + \beta_{\textrm{t}, m} = 1, \ \ \forall m \in \mathcal{M},} \quad\\
	& \textcolor{black}{\text{C}_{13}: 
		\varphi_{\textrm{r}, m}, \varphi_{\textrm{t}, m} \in [0, 2\pi), \ \ \ \forall m \in \mathcal{M},}
	\end{align} 
\end{subequations}
in which the minimum SINR requirements $\gamma^{\min}_{\textrm{r}, i}$ and $\gamma^{\min}_{\textrm{t}, j}$ have to be guaranteed for users $i\in\mathcal{U}_{\textrm{r}}$ and $j\in\mathcal{U}_{\textrm{t}}$ according to $\text{C}_{1}$ and $\text{C}_{2}$, respectively. Likewise, $\text{C}_{3}$ and $\text{C}_{4}$ ensure that the minimum EH requirements $E^{\min}_{\textrm{r}, i}$ and $E^{\min}_{\textrm{t}, j}$ are satisfied for users $i\in\mathcal{U}_{\textrm{r}}$ and $j\in\mathcal{U}_{\textrm{t}}$, respectively. In $\text{C}_{5}$, the transmit power budget of the BS has been limited up to $P_{\mathrm{BS}}^{\mathrm{\max}}$, while $\text{C}_{6}$ enforces the overall energy consumption of the active STAR-RIS to respect a maximum allowable threshold, denoted by $P_{I}^{\max}$. The PS ratio  is adopted within $[0,1]$ for users $i\in\mathcal{U}_{\textrm{r}}$ and $j\in\mathcal{U}_{\textrm{t}}$ according to $\text{C}_{7}$ and $\text{C}_{8}$, respectively. 
The reflection
coefficient of the active STAR-RIS is bounded to $a_{\max}$ in $\text{C}_{9}$.
We have constrained the element selection matrix to be in the binary domain in $\text{C}_{10}$. Constraint $\text{C}_{11}$ indicates that the maximum number of selected elements at the STAR-RIS is limited to $M$. Eventually, constraints $\text{C}_{12}$ and $\text{C}_{13}$ regulate the domain of the amplitude and phase shift of the STAR-RIS, respectively.
\subsection{Convexity Analysis}
\par Analyzing $\mathcal{P}$, one can easily verify that it is non-convex. Specifically, $\text{C}_{1}$ is in quadratic form with respect to (w.r.t.) decision variables 
$ \big \lbrace \lbrace \mathbf{w}_{\textrm{r}, i} \rbrace_{i \in \mathcal{U}_{\textrm{r}}} ,\mathbf{A}, \mathbf{\Theta_{\textrm{r}}},  \mathbf{F} \big \rbrace$ and also fractional w.r.t. 
$ \big \lbrace \rho_{\textrm{r}, i}\rbrace_{i \in \mathcal{U}_{\textrm{r}}}$. Similar observation holds for $\text{C}_{2}$ w.r.t. $ \big \lbrace \lbrace \mathbf{w}_{\textrm{t}, j} \rbrace_{j \in \mathcal{U}_{\textrm{t}}} ,\mathbf{A}, \mathbf{\Theta_{\textrm{t}}},  \mathbf{F} , \lbrace \rho_{\textrm{t}, j} \rbrace_{j \in \mathcal{U}_{\textrm{t}}} \big \rbrace$. 
Regarding (\ref{eq_4}), it can be implied that $\text{C}_{3}$ and $\text{C}_{4}$ are in quadratic form w.r.t. $ \big \lbrace \lbrace \mathbf{w}_{\textrm{r}, i} \rbrace_{i \in \mathcal{U}_{\textrm{r}}} ,\mathbf{A}, \mathbf{\Theta_{\textrm{r}}},  \mathbf{F}, \lbrace \rho_{\textrm{r}, i} \rbrace_{i\in \mathcal{U}_{\textrm{r}}}\big \rbrace$, as well as $ \big \lbrace \lbrace \mathbf{w}_{\textrm{t}, j} \rbrace_{j \in \mathcal{U}_{\textrm{t}}} ,\mathbf{A}, \mathbf{\Theta_{\textrm{t}}},  \mathbf{F} , \lbrace \rho_{\textrm{t}, j} \rbrace_{j \in \mathcal{U}_{\textrm{t}}} \big \rbrace$, respectively. By similar investigations in $\text{C}_{5}$, $\text{C}_{6}$ and also the fractional form of the objective function in $\mathcal{P}$, it is concluded that this optimisation problem is non-convex and highly-coupled. Moreover, the coexistence of discrete and continuous optimisation variables renders this problem in a mixed-integer and non-linear programming (MINLP) category, thereby classifying it as non-deterministic polynomial-time hardness (NP-hard). Consequently, seeking a globally optimal solution through brute-force methods, such as exhaustive search, is typically impractical even for moderately sized systems.
\subsection{Solution Strategy}
\par Recently, model-free DRL-based algorithms have been well applied in the literature to solve non-convex optimisation problems, e.g., \cite{learning1,learning2}. However, in cases where their action space exponentially grows with the scalability of the optimisation variables, their effectiveness cannot be guaranteed. In our case, due to the number of highly-coupled optimisation variables in $\mathcal{P}$, the convergence and overall performance of these algorithms is questionable. To address $\mathcal{P}$ in this paper, we adopt a resource allocation mechanism relying on an alternating optimisation mechanism. This mechanism shrinks the degree of coupling among optimisation variables via problem decomposition technique and exploits both classical convex optimisation methods and DRL-based algorithms in an iterative fashion. Towards that goal, we decompose $\mathcal{P}$ into two sub-problems, namely:
\begin{itemize}
	\item Sub-problem (\textit{i}):  This sub-problem jointly optimises the BS transmit beamforming and PS ratio of users, i.e., $\big ( \lbrace \mathbf{w}_{\textrm{r}, i} \rbrace_{i \in \mathcal{U}_{\textrm{r}}}, \lbrace \mathbf{w}_{t,j} \rbrace_{j \in \mathcal{U}_{\textrm{t}}} \big )$ and $\big ( \lbrace \rho_{\textrm{r}, i} \rbrace_{i \in \mathcal{U}_{\textrm{r}}}, \lbrace \rho_{\textrm{t}, j} \rbrace_{j \in \mathcal{U}_{\textrm{t}}} \big )$. The set of optimisation variables is denoted by $\mathcal{T}_{1} = \big ( \lbrace \mathbf{w}_{\textrm{r}, i} \rbrace_{i \in \mathcal{U}_{\textrm{r}}}, \lbrace \mathbf{w}_{t,j} \rbrace_{j \in \mathcal{U}_{\textrm{t}}}, \lbrace \rho_{\textrm{r}, i} \rbrace_{i \in \mathcal{U}_{\textrm{r}}}, \lbrace \rho_{\textrm{t}, j} \rbrace_{j \in \mathcal{U}_{\textrm{t}}} \big )$. 
	\item Sub-problem (\textit{ii}): This sub-problem with the set of optimisation variables $\mathcal{T}_{2} = \big(\mathbf{A}, \mathbf{\Theta}_{\textrm{r}},  \mathbf{\Theta}_{\textrm{t}}, \mathbf{F}\big)$ corresponds to the joint optimisation of amplification factor matrix $\mathbf{A}$, reflection and re-transmission phase shift matrices of the STAR-RIS, i.e., $\mathbf{\Theta}_{\textrm{r}}$ and $\mathbf{\Theta}_{\textrm{t}}$, as well as its binary element selection matrix $\mathbf{F}$.
\end{itemize} 
By iteratively solving both sub-problems, the overall solution would be obtained, once the convergence is achieved.
\vspace*{-0.85em}
\section{Solution to Sub-problem (\textit{i})}\label{SectionIV}
In this section, we propose a solution based on classical convex optimisation method to address the sub-problem ($i$). This sub-problem is formulated as follows:
\begin{subequations}\label{eq5-edited}
	\begin{align}
	\mathcal{P}_{1}: \ \ &  \max_{\mathcal{T}_{1}}\ \ 
	\dfrac{R(\mathcal{T}_{1})}{P\Big(\lbrace \mathbf{w}_{\textrm{r}, i} \rbrace_{i \in \mathcal{U}_{\textrm{r}}}, \lbrace \mathbf{w}_{\textrm{t}, j} \rbrace_{j \in \mathcal{U}_{\textrm{t}}}\Big)} , \\ 
	&  \mathrm{s.t.} \ \     \text{C}_{1}-\text{C}_{8}.
	\end{align} 
\end{subequations} 
While $\Omega_{\textrm{r},i}$ and $\Omega_{\textrm{t},j}$ are of constant values, let us rewrite $P_{\textrm{r},i}$ and $P_{\textrm{t},j}$ in \eqref{PSI_r} and \eqref{PSI_t} in following form:
\begin{equation}\label{harvested_power}
\begin{split}
& P_{\textrm{r}, i}(\Psi_{\textrm{r}, i}^{\text{NL}}) = b_{\textrm{r}, i} - \frac{1}{a_{\textrm{r}, i}} \ln(\frac{M_{\textrm{r}, i} - \Psi_{\textrm{r}, i}^{\text{NL}}}{\Psi_{\textrm{r}, i}^{\text{NL}}}), \ \ i \in \mathcal{U}_{\textrm{r}},\\
& P_{\textrm{t}, j}(\Psi_{\textrm{t}, j}^{\text{NL}}) = b_{\textrm{t}, j} - \frac{1}{a_{\textrm{t}, j}} \ln(\frac{M_{\textrm{t}, j} - \Psi_{\textrm{t}, j}^{\text{NL}}}{\Psi_{\textrm{t}, j}^{\text{NL}}}), \ \ j \in \mathcal{U}_{\textrm{t}}.
\end{split}
\end{equation}
By doing so, $\text{C}_{3}$ and $\text{C}_{4}$ in $\mathcal{P}$ can be rewritten as
\small
\begin{equation}\label{Edited1}
\begin{split}
&\sum\nolimits_{k \in \mathcal{U}_{\textrm{r}}}
\lvert\bar{\mathbf{h}}^{H}_{\textrm{r}, i} \mathbf{w}_{\textrm{r}, k} \rvert^{2} \!\!+\!\! {\sigma^2_{\textrm{r}, z}}\lVert {\mathbf{h}^{H}_{\textrm{r}, i}} \mathbf{A} \mathbf{F \Theta_{\textrm{r}}}  \rVert^{2} \geq \frac{P_{\textrm{r}, i}(E_{\textrm{r}, i}^{\min})}{ 1-\rho_{\textrm{r}, i}}, \ \ \forall i \in \mathcal{U}_{\textrm{r}}, \nonumber
\end{split}
\end{equation}
\begin{equation}\label{Edited1}
\begin{split}
& \sum\nolimits_{k \in \mathcal{U}_{\textrm{t}}}
\lvert\bar{\mathbf{h}}^{H}_{\textrm{t}, j}\mathbf{w}_{\textrm{t}, k}\rvert^{2} \!\!+\!\! {\sigma^{2}_{\textrm{t}, z}}\lVert {\mathbf{h}^{H}_{\textrm{t}, j}} \mathbf{A} \mathbf{F \Theta_{\textrm{t}}}  \rVert^{2} \geq  \frac{P_{\textrm{t}, j}(E_{\textrm{t}, j}^{\min})}{ 1-\rho_{\textrm{t}, j}},\ \ \forall j \in \mathcal{U}_{\textrm{t}}.
\end{split}
\end{equation}
\normalsize
\par In order to reformulate $\mathcal{P}_{1}$ in a more tractable form, let us recast its optimisation variables as $\mathbf{A F} \mathbf{\Theta}_{\textrm{r}} = \mathbf{\Gamma}_{\textrm{r}}$, $\mathbf{A F} \mathbf{\Theta}_{\textrm{t}} = \mathbf{\Gamma}_{\textrm{t}}$, $\mathbf{W}_{\textrm{r},i} = \mathbf{w}_{\textrm{r},i} \mathbf{w}_{\textrm{r},i}^{H}$, $\mathbf{W}_{\textrm{t}, j} = \mathbf{w}_{\textrm{t}, j} \mathbf{w}_{\textrm{t}, j}^{H}$,  $\bar{\mathbf{H}}_{\textrm{r}, i} = \bar{\mathbf{h}}_{\textrm{r}, i} \bar{\mathbf{h}}_{\textrm{r}, i}^{H}$,  $\bar{\mathbf{H}}_{\textrm{t}, j} = \bar{\mathbf{h}}_{\textrm{t}, j} \bar{\mathbf{h}}_{\textrm{t}, j}^{H}$.
By incorporating these transformations and also simple manipulations, $\mathcal{P}_{1}$ can be reformulated as follows
\begin{equation}\label{eq6}
\begin{split}
& {\mathcal{P}_{1}^{\prime}}: \ \  \max_{\mathcal{T}_{1}}\ \ \quad
\dfrac{R(\mathcal{T}_{1})}{P\Big(\lbrace \mathbf{W}_{\textrm{r}, i} \rbrace_{i \in \mathcal{U}_{\textrm{r}}}, \lbrace \mathbf{W}_{\textrm{t}, j} \rbrace_{j \in \mathcal{U}_{\textrm{t}}}\Big)}   \\
& \mathrm{s.t.} \ \   \small \text{C}_{1}: \cfrac{\text{tr}(\bar{\mathbf{H}}_{\textrm{r}, i} \mathbf{W}_{\textrm{r}, i})}{\gamma^{\min}_{\textrm{r}, i}} - \sum_{k \in \mathcal{U}_{\textrm{r}}, k \neq i} \text{tr}(\bar{\mathbf{H}}_{\textrm{r}, i} \mathbf{W}_{\textrm{r}, k}) \geq \\ & \quad \quad \quad {\sigma^2_{\textrm{r}, z}} \lVert {\mathbf{h}^{H}_{\textrm{r}, i}} \mathbf{\Gamma}_{\textrm{r}} \rVert^{2} + {\sigma^{2}_{\textrm{r}, i}} + \cfrac{{\delta^{2}_{\textrm{r}, i}}}{\rho_{\textrm{r}, i}} , ~~~\forall i\in \mathcal{U}_{\textrm{r}}, \quad \\
&   \small \text{C}_{2}: \cfrac{\text{tr}(\bar{\mathbf{H}}_{\textrm{t}, j} \mathbf{W}_{\textrm{t}, j})}{\gamma^{\min}_{t, j}} - \sum_{k \in \mathcal{U}_{\textrm{t}}, k \neq j} \text{tr}(\bar{\mathbf{H}}_{\textrm{t}, j} \mathbf{W}_{\textrm{t}, k}) \geq \\ & \quad \quad \quad {\sigma^2_{\textrm{t}, z}} \lVert {\mathbf{h}^{H}_{\textrm{t}, j}} \mathbf{\Gamma}_{\textrm{t}} \rVert^{2} + {\sigma^{2}_{\textrm{t}, j}} + \cfrac{{\delta^2_{\textrm{t}, j}}}{\rho_{\textrm{t}, j}} , ~~~\forall j\in \mathcal{U}_{\textrm{t}}, \quad \\ 	&   \text{C}_{3}: \sum_{k \in \mathcal{U}_{\textrm{r}}} \text{tr}(\bar{\mathbf{H}}_{\textrm{r}, i} \mathbf{W}_{\textrm{r}, k}) + {\sigma^2_{\textrm{r}, z}} \lVert {\mathbf{h}^{H}_{\textrm{r}, i}} \mathbf{\Gamma}_{\textrm{r}} \rVert^{2} \geq \cfrac{P_{\textrm{r}, i}(E^{\min}_{\textrm{r} , i})}{1 - \rho_{\textrm{r}, i}}, ~\forall i\in \mathcal{U}_{\textrm{r}}, \quad \\
&   \text{C}_{4}: \small \sum_{k \in \mathcal{U}_{\textrm{t}}} \text{tr}(\bar{\mathbf{H}}_{\textrm{t}, j} \mathbf{W}_{\textrm{t}, k}) + {\sigma^{2}_{\textrm{t}, z}} \lVert {\mathbf{h}^{H}_{\textrm{t}, j}} \mathbf{\Gamma}_{\textrm{t}} \rVert^{2} \geq \cfrac{P_{t,j}(E^{\min}_{\textrm{t} , j})}{1 - \rho_{\textrm{t}, j}}, ~\forall j\in \mathcal{U}_{\textrm{t}}, \quad \\
&    \text{C}_{5}: \sum_{i \in \mathcal{U}_{\textrm{r}}} \text{tr}(\mathbf{W}_{\textrm{r}, i}) + \sum_{j \in \mathcal{U}_{\textrm{t}}} \text{tr}(\mathbf{W}_{\textrm{t}, j}) \leq P_{\mathrm{BS}}^{\max}, \quad \\
&  \small  \text{C}_{6}: \sum_{i \in \mathcal{U}_{\textrm{r}}} \text{tr}( \mathbf{\Gamma}_{\textrm{r}}\mathbf{G} \mathbf{W}_{\textrm{r}, i} \mathbf{G}^{H} \mathbf{\Gamma}_{\textrm{r}}^{H}  ) + \sum_{j \in \mathcal{U}_{\textrm{t}}} \text{tr}( \mathbf{\Gamma}_{\textrm{t}}\mathbf{G} \mathbf{W}_{\textrm{t}, j} \mathbf{G}^{H} \mathbf{\Gamma}_{\textrm{t}}^{H}) \\
& \quad \quad  + {\sigma^2_{\textrm{r}, z}} \lVert \mathbf{\Gamma}_{\textrm{r}} \rVert_{F}^{2} + {\sigma^2_{\textrm{t}, z}} \lVert \mathbf{\Gamma}_{\textrm{t}} \rVert_{F}^{2}\leq P_{I}^{\max}, \quad \\
&  \text{C}_{7}:  0 \le {\rho_{\textrm{r}, i}} \le 1, \quad \quad \forall i \in \mathcal{U}_{\textrm{r}}, \quad  \\
& \text{C}_{8}: 0 \le {\rho_{\textrm{t}, j}} \le 1, \quad \quad \forall j \in \mathcal{U}_{\textrm{t}}, \quad \\
& \text{C}_{9}: \mathbf{W}_{\textrm{r}, i} \succeq 0, \quad \mathbf{W}_{\textrm{t}, j} \succeq 0 \quad \forall i \in \mathcal{U}_{\textrm{r}}, \quad \forall j \in \mathcal{U}_{\textrm{t}}, \\
& \text{C}_{10}: \text{rank}(\mathbf{W}_{\textrm{r}, i}) \leq 1, \quad \text{rank}(\mathbf{W}_{\textrm{t}, j}) \leq 1 \quad \forall i \in \mathcal{U}_{\textrm{r}}, \ \ \forall j \in \mathcal{U}_{\textrm{t}},
\end{split} 
\end{equation}
where $R(\mathcal{T}_{1})$ and $P(\lbrace \mathbf{W}_{\textrm{r}, i} \rbrace_{i \in \mathcal{U}_{\textrm{r}}}, \lbrace \mathbf{W}_{\textrm{t}, j} \rbrace_{j \in \mathcal{U}_{\textrm{t}}})$ are defined as (\ref{eq26a}) and (\ref{eq27_edited}), respectively.
\small
\begin{table*}
	\begin{equation}\label{eq26a}
	\scriptsize \begin{split}
	& R(\mathcal{T}_{1}) =  \sum_{i \in \mathcal{U}_{\textrm{r}}} \Bigg[  \log_{2}  \Big \lbrace  \sum_{k \in \mathcal{U}_{\textrm{r}}}\text{tr}( \bar{\mathbf{H}}_{\textrm{r}, i} \mathbf{W}_{\textrm{r}, k} )  + {\sigma^{2}_{\textrm{r}, z}} \lVert  {\mathbf{h}^{H}_{\textrm{r}, i}} \mathbf{\Gamma_{\textrm{r}}} \rVert^{2} + {\sigma^{2}_{\textrm{r}, i}} + \cfrac{{\delta^{2}_{\textrm{r}, i}}}{\rho_{\textrm{r}, i}}  \Big\rbrace -\log_{2}  \Big \lbrace  \sum_{k \in \mathcal{U}_{\textrm{r}}, k \neq i}\text{tr}( \bar{\mathbf{H}}_{\textrm{r}, i} \mathbf{W}_{\textrm{r}, k} )  + {\sigma^{2}_{\textrm{r}, z}} \lVert  {\mathbf{h}^{H}_{\textrm{r}, i}} \mathbf{\Gamma_{\textrm{r}}} \rVert^{2} + {\sigma^{2}_{\textrm{r}, i}} + \cfrac{{\delta^{2}_{\textrm{r}, i}}}{\rho_{\textrm{r}, i}}  \Big\rbrace \Bigg] \\
	&  +\sum_{j \in \mathcal{U}_{\textrm{t}}} \Bigg[  \log_{2}  \Big \lbrace  \sum_{k \in \mathcal{U}_{\textrm{r}}}\text{tr}( \bar{\mathbf{H}}_{\textrm{t}, j} \mathbf{W}_{\textrm{t}, k} )  + {\sigma^{2}_{\textrm{t}, z}} \lVert  {\mathbf{h}^{H}_{\textrm{t}, j}} \mathbf{\Gamma_{\textrm{t}}} \rVert^{2} + {\sigma^{2}_{\textrm{t}, j}} + \cfrac{{\delta^{2}_{\textrm{t}, j}}}{\rho_{\textrm{t}, j}}  \Big\rbrace -\log_{2}  \Big \lbrace  \sum_{k \in \mathcal{U}_{\textrm{t}}, k \neq j}\text{tr}( \bar{\mathbf{H}}_{\textrm{t}, j} \mathbf{W}_{\textrm{t}, k})  + {\sigma^2_{\textrm{t}, z}} \lVert  {\mathbf{h}^{H}_{\textrm{t}, j}} \mathbf{\Gamma_{\textrm{t}}} \rVert^{2} + {\sigma^{2}_{\textrm{t}, j}} + \cfrac{{\delta^{2}_{\textrm{t}, j}}}{\rho_{\textrm{t}, j}}  \Big\rbrace \Bigg], \\
	\end{split}
	\end{equation}	
	\hrule
\end{table*}
\begin{table*}
	\begin{equation}\label{eq27_edited}
	\scriptsize \begin{split}
	P\Big(\lbrace \mathbf{W}_{\textrm{r}, i} \rbrace_{i \in \mathcal{U}_{\textrm{r}}}, \lbrace \mathbf{W}_{\textrm{t}, j} \rbrace_{j \in \mathcal{U}_{\textrm{t}}}\Big)  =&  \sum_{i \in \mathcal{U}_{\textrm{r}}} \text{tr} (\mathbf{W}_{\textrm{r}, i}) +  \sum_{j \in \mathcal{U}_{\textrm{t}}} \text{tr} (\mathbf{W}_{\textrm{t}, j}) + \zeta  \sum_{i \in \mathcal{U}_{\textrm{r}}} \text{tr}( \mathbf{\Gamma}_{\textrm{r}}\mathbf{G} \mathbf{W}_{\textrm{r}, i} \mathbf{G}^{H} \mathbf{\Gamma}_{\textrm{r}}^{H}  )  +\zeta \sum_{j \in \mathcal{U}_{\textrm{t}}} \text{tr}( \mathbf{\Gamma}_{\textrm{t}}\mathbf{G} \mathbf{W}_{\textrm{t}, j} \mathbf{G}^{H} \mathbf{\Gamma}_{\textrm{t}}^{H}) \\
	& + \zeta {\sigma^{2}_{\textrm{r}, z}} \lVert \mathbf{\Gamma}_{\textrm{r}} \rVert_{F}^{2} + \zeta {\sigma^{2}_{\textrm{t}, z}} \lVert \mathbf{\Gamma}_{\textrm{t}} \rVert_{F}^{2} + M(P_{\text{c}} + P_{\text{DC}}) + P_{\text{BS}}^{\text{Cir}} + \sum_{i \in \mathcal{U}_{\textrm{r}}} p_{i}^{\text{Cir}} + \sum_{j \in \mathcal{U}_{\textrm{t}}} p_{j}^{\text{Cir}},
	\end{split}
	\end{equation}\hrule
\end{table*} 
\normalsize Meanwhile, the rank constraints in $\text{C}_{10}$ are obtained by invoking the SDR method. Now, the sub-problem ${\mathcal{P}_{1}^{\prime}}$ can be reformulated as follows
\begin{subequations}\label{eq7}
	\begin{align}
	\mathcal{P}_{1}^{\prime\prime}: \ \ &   
	\max_{\mathcal{T}_{1}}\ \ 
	\dfrac{R(\mathcal{T}_{1})}{P\Big(\lbrace \mathbf{W}_{\textrm{r}, i} \rbrace_{i \in \mathcal{U}_{\textrm{r}}}, \lbrace \mathbf{W}_{\textrm{t}, j} \rbrace_{j \in \mathcal{U}_{\textrm{t}}}\Big)}   \\ 
	&  \mathrm{s.t.} \ \  \quad \quad \quad \quad \text{C}_{1}-\text{C}_{9}.
	\end{align} 
\end{subequations}
This sub-problem is still non-convex, because of its fractional form and also the non-convex function $R(\mathcal{T}_{1})$ in the numerator of its objective function. We now recast it into a more tractable form as a DC method, which is modeled as
\begin{equation}\label{miani}
\begin{split}
R(\mathcal{T}_{1}) = \ &  f(\mathcal{T}_{1}) - g(\mathcal{T}_{1}),
\end{split}
\end{equation}
\normalsize
in which $f(\mathcal{T}_{1})$ and $g(\mathcal{T}_{1})$ are respectively given in (\ref{f}) and (\ref{g}). 
\small
\begin{table*}
	\begin{equation}\label{f}
	\scriptsize \begin{split}
	f(\mathcal{T}_{1})  = &\sum_{i \in \mathcal{U}_{\textrm{r}}}  \log_{2}  \Big \lbrace  \sum_{k \in \mathcal{U}_{\textrm{r}}}\text{tr}( \bar{\mathbf{H}}_{\textrm{r}, i} \mathbf{W}_{\textrm{r}, k})  + {\sigma^{2}_{\textrm{r}, z}} \lVert  {\mathbf{h}^{H}_{\textrm{r}, i}} \mathbf{\Gamma_{\textrm{r}}} \rVert^{2} + {\sigma^{2}_{\textrm{r}, i}} + \cfrac{{\delta^2_{\textrm{r}, i}}}{\rho_{\textrm{r}, i}}  \Big\rbrace
	+ \sum_{j \in \mathcal{U}_{\textrm{t}}}  \log_{2}  \Big \lbrace  \sum_{k \in \mathcal{U}_{\textrm{r}}}\text{tr}( \bar{\mathbf{H}}_{\textrm{t}, j} \mathbf{W}_{\textrm{t}, k} )  + {\sigma^{2}_{\textrm{t}, z}} \lVert  {\mathbf{h}^{H}_{\textrm{t}, j}} \mathbf{\Gamma_{\textrm{t}}} \rVert^{2} + {\sigma^{2}_{\textrm{t}, j}} + \cfrac{{\delta^{2}_{\textrm{t}, j}}}{\rho_{\textrm{t}, j}}  \Big\rbrace,
	\end{split}
	\end{equation}\hrule
	\scriptsize \begin{equation}\label{g}
	\begin{split}
	g(\mathcal{T}_{1})  = &\sum_{i \in \mathcal{U}_{\textrm{r}}} \log_{2}  \Big \lbrace  \sum_{k \in \mathcal{U}_{\textrm{r}}, k \neq i}\text{tr}( \bar{\mathbf{H}}_{\textrm{r}, i} \mathbf{W}_{\textrm{r}, k})  + {\sigma^2_{\textrm{r}, z}} \lVert  {\mathbf{h}^{H}_{\textrm{r}, i}} \mathbf{\Gamma_{\textrm{r}}} \rVert^{2} + {\sigma^2_{\textrm{r}, i}} + \cfrac{{\delta^{2}_{\textrm{r}, i}}}{\rho_{\textrm{r}, i}}  \Big\rbrace
	+ \sum_{j \in \mathcal{U}_{\textrm{t}}} \log_{2}  \Big \lbrace  \sum_{k \in \mathcal{U}_{\textrm{t}}, k \neq j}\text{tr}( \bar{\mathbf{H}}_{\textrm{t}, j} \mathbf{W}_{\textrm{t}, k} )  + {\sigma^{2}_{\textrm{t}, z}} \lVert  {\mathbf{h}^{H}_{\textrm{t}, j}} \mathbf{\Gamma_{\textrm{t}}} \rVert^{2} + {\sigma^2_{\textrm{t}, j}} + \cfrac{{\delta^2_{\textrm{t}, j}}}{\rho_{\textrm{t}, j}}  \Big\rbrace.
	\end{split}
	\end{equation}
	\hrule
\end{table*}
\begin{table*}
	\begin{equation}\label{g-est}
	\scriptsize\begin{split}
	& g(\mathcal{T}_{1}) \!\leq  \ g(\mathcal{T}_{1}^{(\tau - 1)}) + \sum_{i \in \mathcal{U}_{\textrm{r}}} \text{tr} \Big[   \Big( \nabla_{\mathbf{W}_{\textrm{r}, i}}^{H} g(\mathcal{T}_{1}^{(\tau - 1)}) \Big)  \Big( \mathbf{W}_{\textrm{r}, i} - \mathbf{W}_{\textrm{r}, i}^{(\tau-1)} \Big) \Big] + \sum_{j \in \mathcal{U}_{\textrm{t}}} \text{tr} \Big[   \Big( \nabla_{\mathbf{W}_{\textrm{t}, j}}^{H} g(\mathcal{T}_{1}^{(\tau - 1)}) \Big)  \Big( \mathbf{W}_{\textrm{t}, j} - \mathbf{W}_{\textrm{t}, j}^{(\tau-1)} \Big) \Big] \\
	& + \sum_{i \in \mathcal{U}_{\textrm{r}}} \Big[   \Big( \nabla_{\rho_{\textrm{r}, i}}^{H} g(\mathcal{T}_{1}^{(\tau - 1)}) \Big) \Big( \rho_{\textrm{r}, i} - \rho_{\textrm{r}, i}^{(\tau-1)} \Big) \Big] + \sum_{j \in \mathcal{U}_{\textrm{t}}} \Big[   \Big( \nabla_{\rho_{\textrm{t}, j}}^{H} g(\mathcal{T}_{1}^{(\tau - 1)}) \Big)  \Big( \rho_{\textrm{t}, j} - \rho_{\textrm{t}, j}^{(\tau-1)} \Big) \Big] \approx \widetilde{g}(\mathcal{T}_{1}).
	\end{split}
	\end{equation}
	\hrule
\end{table*}
\normalsize
Note that both $f(\mathcal{T}_{1})$ and $g(\mathcal{T}_{1})$ are convex w.r.t. $\mathcal{T}_{1}$, but $R(\mathcal{T}_{1})$ is not necessarily convex. Here, we leverage the first-order Taylor expansion at point $(\tau)$ to obtain an affine function $\widetilde{g}(\mathcal{T}_{1})$, instead of $g(\mathcal{T}_{1})$. Since, $g(\mathcal{T}_{1})$ is a differentiable
convex function w.r.t. $\lbrace \mathbf{W}_{\textrm{r}, i} \rbrace_{i \in \mathcal{U}_{\textrm{r}}}$, $\lbrace \mathbf{W}_{\textrm{t}, j} \rbrace_{j \in \mathcal{U}_{\textrm{t}}}$, $\lbrace \rho_{\textrm{r}, i} \rbrace_{i \in \mathcal{U}_{\textrm{r}}}$, and $\lbrace \rho_{\textrm{t}, j} \rbrace_{j \in \mathcal{U}_{\textrm{t}}}$, we have 
$\nabla_{\mathbf{W}_{\textrm{r}, i}}^{H} g(\mathcal{T}_{1}^{(\tau - 1)}) = \frac{\mathbf{E}_{i}}{A}$, $\nabla_{\mathbf{W}_{\textrm{t}, j}}^{H} g(\mathcal{T}_{1}^{(\tau - 1)}) = \frac{\mathbf{E}_{j}}{B}$,
$\nabla_{\rho_{\textrm{r}, i}}^{H} g(\mathcal{T}_{1}^{(\tau - 1)}) =  \cfrac{-\frac{{\delta^2_{\textrm{r}, i}}}{\rho_{\textrm{r}, i}^{(\tau-1)}}}{C} $, \normalsize and 
\small $\nabla_{\rho_{\textrm{t}, j}}^{H} g(\mathcal{T}_{1}^{(\tau - 1)}) = \cfrac{-\frac{{\delta^2_{\textrm{t}, j}}}{\rho_{\textrm{t}, j}^{(\tau-1)}}}{D}$, \normalsize where
\small
\begin{equation}
\begin{split}
&A = \sum_{k \in \mathcal{U}_{\textrm{r}}, k \neq i}\text{tr}( \bar{\mathbf{H}}_{\textrm{r}, i} \mathbf{W}_{\textrm{r}, k}^{(\tau -1)} )  + {\sigma^2_{\textrm{r}, z}} \lVert  {\mathbf{h}^{H}_{\textrm{r}, i}} \mathbf{\Gamma_{\textrm{r}}} \rVert^{2} + {\sigma^2_{\textrm{r}, i}} + \cfrac{{\delta^2_{\textrm{r}, i}}}{\rho_{\textrm{r}, i}^{(\tau-1)}}, \nonumber\\
& B = \sum_{k \in \mathcal{U}_{\textrm{t}}, k \neq j}\text{tr}( \bar{\mathbf{H}}_{\textrm{t}, j} \mathbf{W}_{\textrm{t}, k}^{(\tau -1)} )  + {\sigma^2_{\textrm{t}, z}} \lVert  {\mathbf{h}^{H}_{\textrm{t}, j}} \mathbf{\Gamma_{\textrm{t}}} \rVert^{2} + {\sigma^2_{\textrm{t}, j}} + \cfrac{{\delta^2_{\textrm{t}, j}}}{\rho_{\textrm{t}, j}^{(\tau-1)}},\nonumber\\
& C = \\
& \ln 2 \Big(\sum_{k \in \mathcal{U}_{\textrm{r}}, k \neq i}\text{tr}( \bar{\mathbf{H}}_{\textrm{r}, i} \mathbf{W}_{\textrm{r}, k}^{(\tau -1)} )  + {\sigma^2_{\textrm{r}, z}} \lVert  {\mathbf{h}^{H}_{\textrm{r}, i}} \mathbf{\Gamma_{\textrm{r}}} \rVert^{2} + {\sigma^2_{\textrm{r}, i}} + \cfrac{{\delta^2_{\textrm{r}, i}}}{\rho_{\textrm{r}, i}^{(\tau-1)}} \Big), \nonumber \\
& \small D = \\
& \ln 2 \Big(\sum_{k \in \mathcal{U}_{\textrm{t}}, k \neq j}\text{tr}( \bar{\mathbf{H}}_{\textrm{t}, j} \mathbf{W}_{\textrm{t}, k}^{(\tau -1)} )  + {\sigma^2_{\textrm{t}, z}} \lVert  {\mathbf{h}^{H}_{\textrm{t}, j}} \mathbf{\Gamma_{\textrm{t}}} \rVert^{2} + {\sigma^2_{\textrm{t},j}} + \cfrac{{\delta^2_{\textrm{t}, j}}}{\rho_{\textrm{t},j}^{(\tau-1)}}\Big),
 \end{split}
\end{equation}
and
\begin{align}\label{eq32}
\mathbf{E}_{i} &=\left\lbrace \begin{array}{lc}
\frac{\bar{\mathbf{H}}_{\textrm{r}, i}}{\ln 2}, & \mathrm{if} \: \: \: \: i \neq k ,\\
0,&\mathrm{if} \: \: i = k .\\
\end{array}\right. 
&\mathbf{E}_{j} =\left\lbrace \begin{array}{lc}
\frac{\bar{\mathbf{H}}_{\textrm{t},j}}{\ln 2}, & \mathrm{if} \: \: \: \: j \neq k ,\\
0,&\mathrm{if} \: \: j = k .\\
\end{array}\right. 
\end{align}

\normalsize
We note that the difference of  affine and convex functions is always convex and efficiently solvable via existing convex optimisation solvers, e.g., CVX~\cite{es9}, \cite{khalili2019antenna}. So, $R(\mathcal{T}_{1})$ can be approximated as
\begin{equation}\label{miani2}
\begin{split}
\widetilde{R}(\mathcal{T}_{1}) = \ &  f(\mathcal{T}_{1}) 
- \widetilde{g}(\mathcal{T}_{1}),
\end{split}
\end{equation}
which is convex. 
Now, the fractional form of the objective function in $\mathcal{P}_{1}^{\prime\prime}$ is the only reason to its non-convex form. To deal with this issue, we reformulate the objective function on the basis of the Dinkelbach method to obtain a linear form~\cite{es9},~\cite{khalili2019antenna}. Accordingly, the sub-problem $\mathcal{P}_{1}^{\prime\prime}$ can be rewritten as
\begin{subequations}\label{eq34}
	\begin{align}
	\mathcal{P}_{1}^{\prime\prime\prime}: \ \ &  \max_{\mathcal{T}_{1}}\ \ 
	\widetilde{R}((\mathcal{T}_{1}))  - \mu P\Big(\lbrace \mathbf{W}_{\textrm{r}, i} \rbrace_{i \in \mathcal{U}_{\textrm{r}}}, \lbrace \mathbf{W}_{\textrm{t}, j} \rbrace_{j \in \mathcal{U}_{\textrm{t}}}\Big)  \\ 
	& \mathrm{s.t.} \ \  \quad \quad \quad \quad \quad \text{C}_{1}-\text{C}_{9},\\ \nonumber
	\end{align} 
\end{subequations}
\normalsize
in which $\mu = \cfrac{\widetilde{R}(\mathcal{T}_{1})}{P\Big(\lbrace \mathbf{W}_{\textrm{r}, i} \rbrace_{i \in \mathcal{U}_{\textrm{r}}}, \lbrace \mathbf{W}_{\textrm{t}, j} \rbrace_{j \in \mathcal{U}_{\textrm{t}}}\Big)}.$
This sub-problem is now convex and can be directly solved via existing off-the-shelf tools such as CVXPY~\cite{CVXPY}.
\small
\begin{algorithm}[h!]
	\caption{Joint Beamforming and PS Ratio Algorithm.}
	\label{Algorithm 1}
	\begin{algorithmic}[1]
		\State \textbf{Input:} 
		The maximum number of iterations $K_{\max}$, the maximum error tolerance $\epsilon = 10^{-2}$. Define $k=0$ and $\mu^{(k)} = 0.$
		\normalsize \State \textbf{repeat} 
		\State \quad Calculate $\widetilde{g}(\mathcal{T}_{1})$ acording to \eqref{g-est}.
		\State \quad Solve problem $(\mathcal{P}_{1}^{\prime\prime\prime})$ for a given $\mu^{(k)}$ and obtain $\lbrace \mathcal{T}_{1}^{(k)} \rbrace$.
		\State \quad \small \textbf{if} $\Big | \widetilde{R}(\mathcal{T}_{1}^{(k)}) - \mu^{(k)} P\Big(\lbrace \mathbf{W}_{\textrm{r}, i}^{(k)} \rbrace_{i \in \mathcal{U}_{\textrm{r}}}, \lbrace \mathbf{W}_{\textrm{t}, j}^{(k)} \rbrace_{j \in \mathcal{U}_{\textrm{t}}}\Big) \Big | < \epsilon$ then
		\normalsize
		\State \quad \quad \  \textbf{return} $\lbrace \mathcal{T}_{1}^{(*)}\rbrace = \lbrace \mathcal{T}_{1}^{(k)} \rbrace$ and $\mu^{*} = \cfrac{\widetilde{R}(\mathcal{T}_{1}^{(k)})}{P\Big(\lbrace \mathbf{W}_{\textrm{r}, i}^{(k)} \rbrace_{i \in \mathcal{U}_{\textrm{r}}}, \lbrace \mathbf{W}_{\textrm{t}, j}^{(k)} \rbrace_{j \in \mathcal{U}_{\textrm{t}}}\Big)}$.
		\State \quad \small \textbf{else}
		\State \quad \quad \quad \quad \quad $\mu^{(i+1)} = \cfrac{\widetilde{R}(\mathcal{T}_{1}^{(k)})}{P\Big(\lbrace \mathbf{W}_{\textrm{r}, i}^{(k)} \rbrace_{i \in \mathcal{U}_{\textrm{r}}}, \lbrace \mathbf{W}_{\textrm{t}, j}^{(k)} \rbrace_{j \in \mathcal{U}_{\textrm{t}}}\Big)}$.
		\State \quad \textbf{end if}
		\State \quad $k = k + 1.$
		\State \textbf{until} $k = K_{\max}.$
	\end{algorithmic}
\end{algorithm}
\normalsize 

\vspace*{-2em}

\section{Solution to Sub-problem (\textit{ii})}\label{MDS}

\vspace*{-1em}

The second sub-problem optimises the set of optimisation variables $\mathcal{T}_{2}$, under the assumption of optimised set of variables $\mathcal{T}_{1}$ from the previous section. The sub-problem $(\textit{ii})$ is formulated as
\begin{subequations}\label{eq5-edited2}
	\begin{align}
	\mathcal{P}_{2}: \ \ &  \max_{\mathcal{T}_{2}}\ \ 
	\dfrac{R(\mathcal{T}_{2})}{P(\mathcal{T}_{2})}  \\ 
	& \ \ \mathrm{s.t.} \ \   \text{C}_{1}-\text{C}_{13},
	\end{align} 
\end{subequations}
which is non-convex, due to the non-convex form of the objective function and constraints, we elaborated earlier. To address this sub-problem, a novel adaptive solution strategy based on model-free DRL is proposed.
Specifically, we develop our proposed MDS algorithm, relying on two well-known algorithms, namely DDPG and SAC. Towards this goal, we reformulate $\mathcal{P}_2$ in Markov decision process (MDP) form.

\subsection{Reformulation of $\mathcal{P}_2$}
We consider the BS as a central controller that plays the role of DRL agent and interacts with the active STAR-RIS-aided communication system as the environment. The agent performs an action $\mathbf{a}_{\tau}$ in time step $\tau$ based on the state of the environment in this time step, which is denoted by $\mathbf{s}_{\tau}$. By doing so, the environment reconfigures to a new state $\mathbf{s}_{\tau+1}$ and accordingly, a reward function $r(\mathbf{s}_{\tau}, \mathbf{a}_{\tau})$ evaluates the effectiveness of the adopted action. MDP form, in essence, relies on six key elements including action space, state space, reward function, policy, state-action-value function and transition probability. Hence, the MDP form of the sub-problem $\mathcal{P}_2$ can be described as follows. 
\begin{itemize}
	\item \textbf{Action Space:} We define the actions of the DRL agent in conjunction with the set of optimisation variables $\mathcal{T}_{2}$ in sub-problem $\mathcal{P}_{2}$. 
	The action $\mathbf{a}_{\tau}$ in time step $\tau$ is classified as $\mathbf{a}_{\tau} = \lbrace \mathbf{a}_{\tau}^{(1)}, \mathbf{a}_{\tau}^{(2)} \rbrace$, in which the first set corresponds to the STAR-RIS element selection variable with discrete domain, i.e.,  $\mathbf{a}_{\tau}^{(1)} =\lbrace \mathbf{F}_{\tau} \rbrace$. In contrast, the remaining optimisation variables in $\mathcal{T}_2$ with continuous domain are categorized as $\mathbf{a}_{\tau}^{(2)} =\lbrace \mathbf{A}_{\tau}, \mathbf{\Theta}_{\text{r}, \tau}, \mathbf{\Theta}_{\textrm{t}, \tau}\rbrace$. 
	The set of all actions carried out by the agent in all time steps construct the action space $\mathcal{A}$, such that, $\mathbf{a}_{\tau} \in \mathcal{A},\quad\forall \tau$. 
	\item \textbf{State Space:} This space, denoted by $\mathcal{S}$, includes the set of all environment states the agent observes in all time steps. We define a state $\mathbf{s}_{\tau} \in \mathcal{S},\quad\forall \tau,$ to provide the DRL agent with useful information of the environment. The selected state  $\mathbf{s}_{\tau}$ in time step $\tau$ regarding the sub-problem $\mathcal{P}_2$ can be expressed as
	\begin{align} \label{states}
	\mathbf{s}_{\tau}=\lbrace \mathbf{I}_{\tau}, \mathbf{a_{\tau}}, r(\mathbf{s}_\tau, \mathbf{a}_\tau) \rbrace, 
	\end{align}
	in which $	\mathbf{I}_{\tau}=  \big \lbrace \lbrace \gamma_{\text{r},i} \rbrace _{i\in \mathcal{U}_{\text{r}}}, \lbrace \gamma_{\text{t},j} \rbrace _{j\in \mathcal{U}_{\text{t}}},  \lbrace \gamma_{\text{r},i}^ {\text{min}} \rbrace _{i\in \mathcal{U}_{\text{r}}}, \lbrace \gamma_{\text{t},j}^{ \text{min}} \rbrace _{j\in \mathcal{U}_{\text{t}}} \\ ,\lbrace \lbrace E^\text{NL}_{\text{r},i} \rbrace _{i\in \mathcal{U}_{\text{r}}}, E^\text{NL}_{\text{t},j} \rbrace _{j\in \mathcal{U}_{\text{t}}}, \lbrace E_{\text{r},i}^ {\text{min}} \rbrace _{i\in \mathcal{U}_{\text{r}}},  \lbrace E_{\text{t},j}^{\text{min}} \rbrace _{j\in \mathcal{U}_{\text{t}}}, \lbrace \bar{\mathbf{h}}^{H}_{\textrm{r}, i} \rbrace _{i\in \mathcal{U}_{\text{r}}},\\ \lbrace \bar{\mathbf{h}}^{H}_{\textrm{t}, j} \rbrace _{j\in \mathcal{U}_{\text{t}}}, P_\text{BS}^\text{max}, P_I^\text{max} \big \rbrace$.
	
	
	\item \textbf{Reward Function:} This function is defined relying on the objective function in $\mathcal{P}_{2}$ and the satisfaction/violation of its constraints. Accordingly, the reward function in time step $\tau$ evaluates the immediate effectiveness of the action $\mathbf{a}_\tau \in \mathcal{A}$ taken by the DRL agent, when it observes the environment in the state $\mathbf{s}_\tau \in \mathcal{S}$. This function is defined as
	\begin{equation}\label{reward_func}
	\begin{split}
	r(\mathbf{s}_{\tau}, \mathbf{a}_{\tau}) = \text{EE} + \sum_{x=1}^{13} \big(\alpha_{x} \times \text{EE}\big),
	\end{split}
	\end{equation}
	where $\alpha_{x}=0, \ l \in \lbrace 1, \dots, 13 \rbrace$, if the $\text{C}_{x}$-th constraint in $\mathcal{P}_2$ is satisfied and $-1$, otherwise\footnote{Since the constraints $\text{C}_{5}$, $\text{C}_{7}$ and $\text{C}_{8}$ were satisfied in the last section and don't include any optimisation variable from $\mathcal{T}_{2}$, we set $\alpha_5=\alpha_7=\alpha_8=0$.}. Therefore, 
	the reward function would not include any penalty in the case where all constraints are satisfied. 
	\item \textbf{Policy:} The policy $\mu$ is either of deterministic $\mu({\mathbf{s}_\tau})\in \mathcal{A}$ or stochastic $\mu(\mathbf{a}_\tau | {\mathbf{s}_\tau})$ type. In the former case, a mapping is characterized between the state of the environment and the action the DRL agent adopts. In contrast, the latter case introduces the probability, with which the DRL agent takes the action $\mathbf{a}_\tau \in \mathcal{A}$ in time step $\tau$, while observing the environment in the state ${\mathbf{s}_\tau} \in \mathcal{S}$. 
	\item \textbf{State-action-value function:} The long-term effectiveness of the adopted policy $\mu$, whether of deterministic or stochastic type, is evaluated using the Q-value function or equivalently, state-action-value function declared by $q_{\mu}(\mathbf{s}_{\tau}, \mathbf{a}_{\tau})$. This function, unlike the immediate reward function, offers a cumulative long-term reward between consecutive states of the environment for the DRL agent.
\end{itemize}
\vspace*{-1em}
\subsection{Modified-DDPG}
The proposed modified-DDPG algorithm consists of four neural networks, i.e., an actor network (AN) with parameters $\boldsymbol{\theta}$, a critic network (CN) with parameters $\boldsymbol{\phi}$, a target actor network (TAN) with parameters ${\boldsymbol{\bar {\theta}}}$, and a target critic network (TCN) with parameters ${\boldsymbol{\bar \phi}}$. 
More specifically, the AN takes a state $\mathbf{s}_\tau$ as an input and outputs the action $\mathbf{a}_\tau^{(1)}$. 
Then, the CN takes the action $\mathbf{a}_\tau^{(1)}$ and state $\mathbf{s}_\tau$ as input and outputs ${q_\mu}( {\mathbf{s}_\tau, \mathbf{a}_\tau^{(1)}; \boldsymbol{\phi}})$, which is a Q-value by taking action $\mathbf{a}_\tau^{(1)}$ in state $\mathbf{s}_\tau$ under policy $\mu$. Furthermore, the TAN takes the state $\mathbf{s}_{\tau+1}$ as input and outputs the action $\mathbf{a}_{\tau+1}^{(1)}$. After receiving
the action vector $\mathbf{a}_{\tau+1}^{(1)}$, the $\tanh$ function is applied to restrict the
actions within the range of $-1$ to $1$. As mentioned before,
the modified-DDPG algorithm is designed for an action space $\mathbf{a}_{\tau+1}^{(1)}$ with values $0$ and $1$, such that if $[a_{\tau+1}^{(1)}]_{i} > 0$, $[a_{\tau+1}^{(1)}]_{i}  = 0 $, and if $[a_{\tau+1}^{(1)}]_{i} < 0$, then $[a_{\tau+1}^{(1)}]_{i} = -1 $ for $i \in |\mathbf{a}_{\tau+1}^{(1)}|$. The TCN takes the action $\mathbf{a}_{\tau+1}^{(1)}$ and the state $\mathbf{s}_{\tau+1}$ as input and outputs ${q_\mu}( {\mathbf{s}_\tau}, \mathbf{a}_\tau^{(1)}; \boldsymbol{\bar \phi} )$, which is a Q-value by taking action $\mathbf{a}_{\tau+1}^{(1)}$ in state $\mathbf{s}_{\tau+1}$ under policy $\mu$. In modified-DDPG algorithm, a replay buffer $\mathcal{B}$ is applied to enable an agent to interact with its environment, such that the $l$-th action $\mathbf{a}_{\tau,l}^{(1)}$ is taken by the agent under the $l$-th state $\mathbf{s}_{\tau,l}$  in time step $\tau$. Then, a reward $ r( {\mathbf{s}_{\tau,l},\mathbf{a}_{\tau,l}^{(1)}})$ is obtained by an agent and a transition to the next step $\mathbf{s}_{\tau+1,l}$ is completed. Hence, a transition tuple $D_B$ $\big(\mathbf{s}_{\tau,l}, \mathbf{a}_{\tau,l}^{(1)}, r( {\mathbf{s}_{\tau,l},\mathbf{a}_\tau^{(1)}}), \mathbf{s}_{\tau+1,l}\big)$ is stored into the replay buffer $\mathcal{B}$. The parameters of the neural networks used in this algorithm must be adjusted using the transition tuple. 

For training of the CN, it is assumed that a mini-batch comprising of ${D_B}$ transition tuples is sampled from the replay buffer $\mathcal{B}$. Then, the target Q value of $l$-th transition tuple $\big(\mathbf{s}_{\tau,l}, \mathbf{a}_{\tau,l}^{(1)}, r( {\mathbf{s}_{\tau,l},\mathbf{a}_\tau^{(1)}}), \mathbf{s}_{\tau+1,l}\big)$ can be expressed as
\small
\begin{align}
{y_{\text{tar}}}(\mathbf{s}_{\tau,l},\mathbf{a}_{\tau,l}^{(1)}) = r\left( {\mathbf{s}_{\tau,l},\mathbf{a}_{\tau,l}^{(1)}} \right) + \xi {q_\mu }\left( {s_{\tau+1,l},\mathbf{a}_{\tau+1,l}^{(1)};\boldsymbol{\bar \phi}} \right),
\end{align}
\normalsize
where $\xi \in (0, 1]$ denotes the discount rate and ${q_\mu }\left( {s_{\tau + 1,l},\mathbf{a}_{\tau+ 1,l}^{(1)};\boldsymbol{\bar \phi}} \right)$ represents a Q-value of next action $\mathbf{a}_{\tau + 1,l}^{(1)}$ under state $\mathbf{s}_{\tau + 1,l}$.
Then, the loss function which is the mean-squared Bellman error (MSBE) between the target Q-value and the Q-value output from the considered CN with parameters $\boldsymbol{\phi}$ is given by
\begin{align}\label{CN-DDPG1}
L\left( \boldsymbol{\phi}  \right) = \frac{1}{{{D_B}}}\sum\limits_{l = 1}^{{D_B}} {\left( {{y_{\text{tar}}}(\mathbf{s}_{\tau,l},\mathbf{a}_{\tau,l}^{(1)}) - {q_\mu }\left({\mathbf{s}_{\tau,l},\mathbf{a}_{\tau,l}^{(1)}; \boldsymbol{\phi}}\right)} \right)},
\end{align}
where $\mathbf{a}_{\tau,l}^{(1)} = \mu \left( {\mathbf{s}_{\tau,l};\boldsymbol{\theta}} \right)$. 
Finally, gradient descent is applied to update the parameters of the CN denoted as $\boldsymbol{\phi} = \boldsymbol{\phi}- \vartheta {\nabla _{\boldsymbol{\phi}}}L\left(\boldsymbol{\phi}\right)$, 
where $0<\vartheta<1$ represents the learning rate.
The parameters of the AN, denoted as $\boldsymbol{\theta}$, are updated to enhance the Q-value by encouraging the action output from the AN in a direction to increase the Q-value. Therefore, a gradient ascent is used to update the AN parameters as follows 
\begin{align}\label{AN-DDPG1}
\boldsymbol{\theta}  = \boldsymbol{\theta}- \vartheta' \frac{1}{{{D_B}}}\sum\limits_{i = 1}^{{D_B}} {{\nabla _{\mu \left( {\mathbf{s}_{\tau,l};\boldsymbol{\theta}} \right)}}} {q_\mu }\left( {\mathbf{s}_{\tau,l},\mathbf{a}_{\tau,l}^{(1)};\boldsymbol{\phi}} \right){\nabla _{\boldsymbol{\theta}}}\mu \left( {\mathbf{s}_{\tau,l};\boldsymbol{\theta}} \right),
\end{align} 
where $0 \le \vartheta' \le 1$ denotes the learning rate. It is worth mentioning that the parameters of TAN and TCN are updated using the soft update method~\cite{DDPG}.
\vspace*{-2em}
\subsection{SAC Algorithm}
SAC algorithm is comprised of five deep neural networks; two critic networks including a critic network $1$ (CN1) with parameter $\boldsymbol{\omega}_1$ and a critic network $2$ (CN2) with parameter $\boldsymbol{\omega}_2$; two target critic networks including a target critic network $1$ (TCN$1$) with parameter $\bar{\boldsymbol{\omega}}_1$ and a target critic network $2$ (TCN$2$) with parameter $\bar{\boldsymbol{\omega}}_2$; and the actor network (AN) with parameter $\boldsymbol{\chi}$. It is worth pointing out that the structure of these two CNs and two TCNs are similar to the structure of the critic network, and the structure of the actor network is the same as the structure of the actor network in the modified-DDPG section. 

The loss function for updating the parameters CN1 can be expressed as
\begin{equation}\label{CN1-SAC1}
\mathcal{L}(\boldsymbol{\omega}_{1}) = \frac{1}{\lvert D_B \rvert} \sum_{l=1}^{D_B} \big( y_{\tau+1,l} - q_{\mu} (\mathbf{s}_{\tau,l}, \mathbf{a}_{\tau,l}^{(2)}; \boldsymbol{\omega}_{1})  \big)^{2},
\end{equation}
where
\begin{equation}
\begin{split}
y_{\tau+1,l} =& r\left( {\mathbf{s}_{\tau,l},\mathbf{a}_{\tau,l}^{(2)}} \right) + \gamma \Big[ \underset{j=1, 2}{\mathrm{min}} \quad q_{\pi}(\mathbf{s}_{\tau+1,l}, \mathbf{a}_{\tau+1,l}^{(2)}, \bar{\boldsymbol{\omega}}_{j}) \\&
-\lambda \log \Big( \pi \big( \mathbf{a}_{\tau+1,l} \big | \mathbf{s}_{\tau+1,l}; \boldsymbol{\chi}\big) \Big)\Big].
\end{split}
\end{equation}
Likewise, the loss function of CN2 is given by
\begin{equation}\label{CN2-SAC1}
\mathcal{L}(\boldsymbol{\omega}_{2}) = \frac{1}{\lvert D_B \rvert} \sum_{l=1}^{D_B} \big( y_{\tau+1,l} - q_{\mu} (\mathbf{s}_{\tau,l}, \mathbf{a}_{\tau,l}^{(2)}; \boldsymbol{\omega}_{2})\big)^{2},
\end{equation}
The loss functions $\mathcal{L}(\boldsymbol{\omega}_{1})$ and $\mathcal{L}(\boldsymbol{\omega}_{1})$ can be minimized employing the gradient descent technique, which involves updating $\boldsymbol{\omega}_{1}$ and $\boldsymbol{\omega}_{2}$. 
Also, the loss function for updating the parameters of AN is
\begin{equation}\label{AN-SAC1}
\begin{split}
\mathcal{L}(\boldsymbol{\chi}) =& \frac{1}{\lvert D_B \rvert} \sum_{l=1}^{D_B} \Big[ \underset{i=1, 2}{\mathrm{min}} \ q_{\mu}(\mathbf{s}_{\tau,l}, \mu_{\boldsymbol{\chi}}(\mathbf{s}_{\tau,l}) + \tilde{\epsilon}_{\tau,l} \sigma_{\boldsymbol{\chi}}(\mathbf{s}_{\tau,l}), \boldsymbol{\omega}_{i}) \\
& -\lambda \log \mu_{\boldsymbol{\chi}}\big(\mu_{\boldsymbol{\chi}}(\mathbf{s}_{\tau,l}) + \tilde{\epsilon}_{\tau,l} \sigma_{\boldsymbol{\chi}}(\mathbf{s}_{\tau,l}) | \mathbf{s}_{\tau,l}; \boldsymbol{\chi}\big)\Big].
\end{split}
\end{equation}
\vspace{-0.2em}
Finally, the parameters of AN are updated using the gradient descent method as $\boldsymbol{\chi} = \boldsymbol{\chi} - \epsilon_{3} \nabla_{\boldsymbol{\chi}} \mathcal{L}(\boldsymbol{\chi})$, 
where $\epsilon_{3}$ is the learning rate. It is noteworthy that the parameters for TCN1 and TCN2 undergo updates faciliated by the soft update method~\cite{SAC}.

The algorithm MDS is a combination of two algorithms, modified-DDPG and SAC, used to handle the continuous optimisation variables $\mathcal{T}_{2}$. As mentioned earlier, the discrete optimisation variables $\mathcal{T}_{1}$ are solved using the classical optimisation approach outlined in Algorithm $1$. We list the pseudo-code of the proposed algorithm, proposed scheme $1$, which is a combination of Algorithm $1$ and the MDS algorithm, in Algorithm $2$.

\small		
\begin{algorithm}[h!]
	\caption{The Proposed Scheme $1$ Algorithm.}
	\label{prposed schem 1}
	\begin{algorithmic}[1]
		\State \textbf{Input:} 
		Number of time slots $T$, number of episodes $E_\text{max}$, and batch size $D_B$.
		\normalsize \State \textbf{Initialization:} Randomly initialize $\boldsymbol{\theta}$ and $\boldsymbol{\phi}$ for modified-DDPG algorithm, as well as $\boldsymbol{\chi}$, $\boldsymbol{\omega}_{1}$, and $\boldsymbol{\omega}_{2}$ for the SAC algorithm. Initialize TAN and TCN for DDPG algorithm with $\bar{\boldsymbol{\theta}} \leftarrow \boldsymbol{\theta}$ and $\bar{\boldsymbol{\phi}} \leftarrow \boldsymbol{\phi}$, as well as TCN1 and TCN2 for SAC algorithm with $\bar{\boldsymbol{\omega}}_1\leftarrow \boldsymbol{\omega}$ and $\bar{\boldsymbol{\omega}}_2 \leftarrow \boldsymbol{\omega}$, initialize the empty replay buffer $\mathcal{B}$.
		\State Initialize optimisation variables in $\mathcal{T}_{2}$.
		\For{\texttt{$E = 1, 2, \dots, E_\text{max}$}}
		\State Solve problem $(\mathcal{P}1^{\prime\prime\prime})$ and get optimisation variables in $\mathcal{T}_{1}^{(\star)}$ by using Algorithm $1$, where optimisation variables $\mathcal{T}_{2}$ is given.
		\State Reset the environment to get the initial state.
		\State The inner $\textbf{for}$ loop solves problem $\mathcal{P}2$ by being provided with the optimisation variables $\mathcal{T}_{1}$, and its output comprises the optimisation variables $\mathcal{T}_{2}$. 
		\For{\texttt{$\tau = 1, 2, \dots, T$}}
		\State Observe state $\mathbf{s}_{\tau}$ and select the discrete action $\mathbf{a}_{\tau,l}^{(1)}$ and the continuous action $\mathbf{a}_{\tau,l}^{(2)}$.
		\State Execute actions $\mathbf{a}_{\tau}^{(1)}$ and $\mathbf{a}_{\tau}^{(2)}$ on the environment.
		\State Observe reward $r({\mathbf{s}_{\tau}, \mathbf{a}_{\tau}^{(1)}, \mathbf{a}_{\tau}^{(2)}})$ and transit to the next state $\mathbf{s}_{\tau+1}$.
		\State Store $(\mathbf{s}_{\tau}, \mathbf{a}_{\tau}^{(1)}, \mathbf{a}_{\tau}^{(2)}, r({\mathbf{s}_{\tau}, \mathbf{a}_{\tau}^{(1)}, \mathbf{a}_{\tau}^{(2)}}), \mathbf{s}_{\tau+1})$ in $\mathcal{B}$.
		\State {Randomly sample a batch set $D_B$ from $\mathcal{B}$.}
		\State \small Update the network parameters $\boldsymbol{\theta}$, $\boldsymbol{\phi}$, $\boldsymbol{\chi}$, $\boldsymbol{\omega}_1$, and $\boldsymbol{\omega}_2$.
		\State Update the target network parameters $\boldsymbol{\bar \theta}$, $\boldsymbol{\bar \phi}$, $\boldsymbol{\bar \omega}_1$, and $\boldsymbol{\bar \omega}_2$.
		\EndFor
		\State Preparation of the optimisation variables $\mathcal{T}_{2}$.
		\EndFor
	\end{algorithmic}
\end{algorithm}
\normalsize
\section{Proposed Meta-Learning Framework}\label{MMDS}
In this section, a meta-learning algorithm is proposed to apply a fast adaptability between training and testing environments in a downlink multi-user scenario with the assistance of an active STAR-RIS. Hence, To take advantage of meta-learning, specifically the
model-agnostic meta-learning
(MAML) approach proposed in~\cite{MAML}, the MMDS algorithm is introduced to integrate meta-learning with modified-DDPG and SAC algorithms for solving problem $\mathcal{P}$. In papers~\cite{meta_V2X},~\cite{faramarzi_meta}, and \cite{AARIS_paper}, meta-learning has been employed in wireless communication, demonstrating that the algorithm's convergence rate and generalization capability are enhanced when the training environment differs from the testing environment. In the MMDS algorithm, a meta-task set $\mathcal{Z}=\{1, \dots , Z\}$ comprising of $Z$ different tasks is considered, while there is only one task in the MDS algorithm. Furthermore, each task is defined as an MDP and a replay buffer for each task is denoted by $\mathcal{B}_{z}$, $z \in \mathcal{Z}$. More specifically, the number of times in which the users are randomly being placed in an initial location determines the number of tasks in problem $\mathcal{P}$. 
Two main stages are considered in the proposed MMDS algorithm, i.e., \textit{meta-training} and \textit{meta-adaptation}. 
In the meta-training stage, the support set denoted by $\mathcal{B}_{z}^{\mathrm{trn}}$ is randomly sampled from the $\mathcal{B}_{z}$ to update the network weights of the actor and critic networks of the modified-DDPG and SAC for each task $z$. Furthermore, the set $\mathcal{B}_{z}^{\mathrm{val}}$ is applied to update the weights of the global network over all tasks $Z$, in which the environment for each task $z$ is reset to obtain the initial state $\mathbf{s}^z_{\tau}$. Next, the agent selects an action $\mathbf{a}^z_{\tau}$ at each time step $\tau$, receives reward $r(\mathbf{s}^z_{\tau}, \mathbf{a}^z_{\tau})$, and transits to the next state $\mathbf{s}^z_{\tau+1}$. The experienced transition $(\mathbf{s}^z_\tau, \mathbf{a}^z_\tau, r\left( {\mathbf{s}^z_\tau,\mathbf{a}^z_\tau} \right), \mathbf{s}^z_{\tau+1})$ is stored in the replay buffer $\mathcal{B}_z$ for each task $z$. Then, a batch set $\mathcal{B}_{z}^{\mathrm{trn}}$ is randomly sampled from $\mathcal{B}_{z}$ in the case the number of experienced transitions overflow the batch size $\mathcal{B}_{z}$. Afterwards, the parameters of the actor and critic networks in modified-DDPG and SAC for each task $z$, respectively denoted by $\hat{\boldsymbol{\theta}}_{z}$,  $\hat{\boldsymbol{\phi}}_{ z}$, $\hat{\boldsymbol{\chi}}_{z}$,  $\hat{\boldsymbol{\omega}}_{1, z}$, and $\hat{\boldsymbol{\omega}}_{2,z}$ are updated as follows 

\small
\begin{align}\label{task_MDDPG}
&\left\lbrace \begin{array}{lc}
\hat{\boldsymbol{\theta}}_{z} = \mathrm{arg} \ \underset{\theta}{\mathrm{min}} \quad \mathcal{L}_{z}(\mathbf{\boldsymbol{\theta}}, \mathcal{B}_z^{\mathrm{trn}} ),\\
\hat{\boldsymbol{\phi}}_{z} = \mathrm{arg} \ \underset{\phi}{\mathrm{min}} \quad \mathcal{L}_{z}(\mathbf{\boldsymbol{\phi}}, \mathcal{B}_z^{\mathrm{trn}}),\ 
\end{array}\right. 
\end{align}
\normalsize
and
\small
\begin{align}\label{task_SAC}
&\left\lbrace \begin{array}{lc}
\hat{\boldsymbol{\chi}}_{z} = \mathrm{arg} \ \underset{\boldsymbol{\chi}}{\mathrm{min}} \quad \mathcal{L}_{z}(\mathbf{\boldsymbol{\chi}}, \mathcal{B}_z^{\mathrm{trn}} ),\\
\hat{\boldsymbol{\omega}}_{1,z}= \mathrm{arg} \ \underset{\boldsymbol{\omega}_{1}}{\mathrm{min}} \quad \mathcal{L}_{z}(\mathbf{\boldsymbol{\omega}}_{1}, \mathcal{B}_z^{\mathrm{trn}} ) ,\\
\hat{\boldsymbol{\omega}}_{2,z} = \mathrm{arg} \ \underset{\boldsymbol{\omega}_{2}}{\mathrm{min}} \quad \mathcal{L}_{z}(\mathbf{\boldsymbol{\omega}}_{2}, \mathcal{B}_z^{\mathrm{trn}}),
\end{array}\right. 
\end{align}
\normalsize
where $\mathcal{L}_{z}(\mathbf{\boldsymbol{\theta}}, \mathcal{B}_z^{\mathrm{trn}})$ and $\mathcal{L}_{z}(\mathbf{\boldsymbol{\phi}}, \mathcal{B}_z^{\mathrm{trn}})$ denote the loss functions of AN and CN in modified-DDPG algorithm for each task $z$, respectively. Moreover, the loss functions of AN, CN1, and CN2 in SAC algorithm for each tast $z$ are respectively given by $\mathcal{L}_{z}(\mathbf{\boldsymbol{\chi}}, \mathcal{B}_z^{\mathrm{trn}})$, $\mathcal{L}_{z}(\mathbf{\boldsymbol{\omega}}_1, \mathcal{B}_z^{\mathrm{trn}})$, and $\mathcal{L}_{z}(\mathbf{\boldsymbol{\omega}}_2, \mathcal{B}_z^{\mathrm{trn}})$. 
In the next step, the evaluation step is performed, where a batch set $\mathcal{B}_{z}^{\mathrm{val}}$ from $\mathcal{B}_{z}$ is sampled. The global parameters including the AN ($\boldsymbol{\theta}$) and CN  ($\boldsymbol{\phi}$) in modified-DDPG algorithm, as well as AN ($\boldsymbol{\chi}$), CN1 ($\boldsymbol{\omega}_1$), and CN2  ($\boldsymbol{\omega}_2$) in SAC algorithm are updated by employing the derivative of their corresponding loss functions. Hence, the optimisation problem global parameters are optimised as follows

\small
\begin{align}\label{global-MDDPG}
&\left\lbrace \begin{array}{lc}
{\boldsymbol{\theta}} = \mathrm{arg} \ \underset{\theta}{\mathrm{min}} \quad \sum_{z} \mathcal{L}_{z} (\hat{\boldsymbol{\theta}}_{z}, \mathcal{B}_{z}^{\mathrm{val}}),\\
{\boldsymbol{\phi}} = \mathrm{arg} \ \underset{\phi}{\mathrm{min}} \quad \sum_{z} \mathcal{L}_{z} (\hat{\boldsymbol{\phi}}_{z}, \mathcal{B}_{z}^{\mathrm{val}}),\ 
\end{array}\right. 
\end{align}
\normalsize
and
\small
\begin{align}\label{global_SAC}
&\left\lbrace \begin{array}{lc}
\mathbf{\boldsymbol{\chi}} = \mathrm{arg} \ \underset{{\chi}}{\mathrm{min}} \quad \sum_{z} \mathcal{L}_{z} (\hat{\boldsymbol{\chi}}_{z}, \mathcal{B}_{z}^{\mathrm{val}}),\\
\boldsymbol{\omega}_{1} = \mathrm{arg} \ \underset{\omega_{1}}{\mathrm{min}} \quad \sum_{z} \mathcal{L}_{z} (\hat{\boldsymbol{\omega}}_{1,z}, \mathcal{B}_{z}^{\mathrm{val}}),\\
\boldsymbol{\omega}_{2} = \mathrm{arg} \ \underset{\omega_{2}}{\mathrm{min}}\quad \sum_{z} \mathcal{L}_{z} (\hat{\boldsymbol{\omega}}_{2,z}, \mathcal{B}_{z}^{\mathrm{val}}).
\end{array}\right. 
\end{align}
\normalsize
\begin{algorithm}[h!]
	\caption{The Meta-Training Phase of the Proposed Scheme 2 Algorithm.}
	\label{meta-training-phase}
	\begin{algorithmic}[1]
		\State \textbf{Input:} 
		Number of time slots $T$, number of tasks $Z$, number of episodes for meta-training phase $E_{\mathrm{trn}}$.
		\State \textbf{Initialization:}
		Randomly initialize $\boldsymbol{\theta}$ and $\boldsymbol{\phi}$ for modified-DDPG algorithm, as well as $\boldsymbol{\chi}$, $\boldsymbol{\omega}_{1}$, and $\boldsymbol{\omega}_{2}$ for the SAC algorithm. Initialize TAN and TCN in modified-DDPG with $\bar{\boldsymbol{\theta}} \leftarrow \boldsymbol{\theta}$ and $\bar{\boldsymbol{\phi}} \leftarrow \boldsymbol{\phi}$, as well as TCN1 and TCN2 in SAC algorithm with $\bar{\boldsymbol{\omega}}_1\leftarrow \boldsymbol{\omega}$ and $\bar{\boldsymbol{\omega}}_2 \leftarrow \boldsymbol{\omega}$. 
		Initialize empty experience memory $\mathcal{B}_{z}$ for each task $ z \in\mathcal{Z} $. Initialize optimisation variables in $\mathcal{T}_{2}$.
		\For{\texttt{$E = 1, 2, \dots, E_{\mathrm{trn}}$}}
		\State Similar to lines 5 to 7 of Algorithm \ref{prposed schem 1}.
		\For{\texttt{$\tau = 1, 2, \dots, T$}}
		\For{\texttt{$z = 1, 2, \dots, Z$}}
		\State Observe state $\mathbf{s}^z_{\tau}$ and select action  $\mathbf{a}_{\tau}^{z}$, execute action $\mathbf{a}_{\tau}^{z}$ on the environment.
		\State Observe next state $\mathbf{s}_{\tau+1}^{z}$ and receive reward $r(\mathbf{s}^z_{\tau}, \mathbf{a}^z_{\tau})$.
		\State Store $\left(\mathbf{s}_\tau^{z}, \mathbf{a}_\tau^{z}, r_\tau^{z}, \mathbf{s}_{\tau+1}^{z} \right)$ in $\mathcal{B}_{z}$.
		\State {Randomly sample a batch set $\mathcal{B}_{z}^{\mathrm{trn}}$ from $\mathcal{B}_{z}$.}
		\State Update network parameters $\hat{\boldsymbol{\theta}}_{z}$ and $\hat{\boldsymbol{\phi}}_{ z}$ for modified-DDPG algorithm by \eqref{task_MDDPG}.
		\State Update network parameters $\hat{\boldsymbol{\chi}}_{z}$,  $\hat{\boldsymbol{\omega}}_{1, z}$, and $\hat{\boldsymbol{\omega}}_{2,z}$  for SAC algorithm by \eqref{task_SAC}.
		\EndFor
		\State Randomly sample a batch set $\mathcal{B}_{z}^{\mathrm{val}}$ from $\mathcal{B}_{z}$ and Calculate the gradient of loss functions $\sum_{z} \mathcal{L}_{z} (\hat{\boldsymbol{\theta}}_{z}, \mathcal{B}_{z}^{\mathrm{val}})$, $\sum_{z} \mathcal{L}_{z} (\hat{\boldsymbol{\phi}}_{z}, \mathcal{B}_{z}^{\mathrm{val}}) $, $\sum_{z} \mathcal{L}_{z} (\hat{\boldsymbol{\chi}}_{z}, \mathcal{B}_{z}^{\mathrm{val}}) $, $\sum_{z} \mathcal{L}_{z} (\hat{\boldsymbol{\omega}}_{1, z}, \mathcal{B}_{z}^{\mathrm{val}})$, and $\sum_{z} \mathcal{L}_{z} (\hat{\boldsymbol{\omega}}_{2,z}, \mathcal{B}_{z}^{\mathrm{val}}) $ for updating $\boldsymbol{\theta}$, $\boldsymbol{\phi}$,  $\boldsymbol{\chi}$,  $\boldsymbol{\omega}_1$, and $\boldsymbol{\omega}_2$ in \eqref{global-MDDPG} and \eqref{global_SAC} using gradient descent.
		\normalsize \EndFor
		\State \small Update the global model parameters $\boldsymbol{\theta}$, $\boldsymbol{\phi}$,  $\boldsymbol{\chi}$,  $\boldsymbol{\omega}_1$, and $\boldsymbol{\omega}_2$.
		\State Preparation of the optimisation variables $\mathcal{T}_{2}$.
		\EndFor
	\end{algorithmic}
\end{algorithm}
\begin{algorithm}[h!]
	\caption{Proposed Scheme 2 Algorithm.}
	\label{meta-adaptation-phase}
	\begin{algorithmic}[1]
		\State \textbf{Input:} 
		Number of time slots $T$, number of episodes for meta-adaptation phase $E_{\mathrm{adp}}$.
		\State \textbf{Initialization:}
		Initialize empty experience memory $\mathcal{B}^{\mathrm{ada}}$. Initialize optimisation variables in $\mathcal{T}_{2}$.
		\algrule
		\begin{center}
			\textbf{Meta-adaptation phase}
		\end{center}
		\State 	Initialize $\mathbf{\Delta} \leftarrow \boldsymbol{\theta}$, $\mathbf{\Omega} \leftarrow \boldsymbol{\phi}$ for modified-DDPG algorithm.
		\State Initialize $\mathbf{\Xi}\leftarrow \boldsymbol{\chi}$, $\mathbf{\Pi}_{1} \leftarrow \boldsymbol{\omega}_{1}$, and $\mathbf{\Pi}_{2} \leftarrow \boldsymbol{\omega}_{2}$ for SAC algorithm.
		\For{$E = 1, 2, \dots, E_{\mathrm{adp}}$}
		\State Similar to lines 5 to 7 of Algorithm \ref{prposed schem 1}.
		\For{$\tau = 1, 2, \dots, T$}			
		\State Generate  experienced transitions $\left(\mathbf{s}_\tau, \mathbf{a}_\tau, r_\tau, \mathbf{s}_{\tau+1} \right)$ similar to steps $7$ and $8$ in the Algorithm \ref{meta-training-phase}.
		\State Store the experienced transitions in $\mathcal{B}^{\mathrm{ada}}$ and then randomly sample a batch set.
		\State Update the network parameters of the modified-DDPG and SAC algorithms $\mathbf{\Delta}$, $\mathbf{\Omega}$, and $\mathbf{\Xi}$, $\mathbf{\Pi}_{1}$, and $\mathbf{\Pi}_{2}$.
		\EndFor
		\State Preparation of the optimisation variables $\mathcal{T}_{2}$.
		\EndFor
	\end{algorithmic}
\end{algorithm}
\normalsize
In the meta-training phase, a learning model for all meta-tasks is trained. Furthermore, in the meta-adaptation phase, the parameters of the model are initialized using the obtained parameters from the Meta-learning rather than a random initialization. In doing so, an experience memory denoted by $\mathcal{B}^\text{adp}$ is utilized to store the experienced transitions consisting of the current state, the chosen action, the received reward of the new agent and the next state. The pseudo-code for meta-training phase of the combination of Algorithm \ref{meta-training-phase} and the proposed MMDS algorithm is outlined in Algorithm \ref{prposed schem 1}. Similar to the meta-training stage, the network parameters including the AN ($\mathbf{\Delta}$) and the CN ($\mathbf{\Omega}$) in DDPG algorithm as well as AN ($\mathbf{\Xi}$), CN1 ($\mathbf{\Pi}_1$), and CN2 ($\mathbf{\Pi}_2$) in the SAC algorithm are utilized. In the training phase, the network parameters are first replaced by the corresponding parameters of the learning model, i.e., $\mathbf{\Delta} \leftarrow \boldsymbol{\theta}$ and $\mathbf{\Omega} \leftarrow \boldsymbol{\phi}$ for modified-DDPG as well as  $\mathbf{\Xi} \leftarrow \boldsymbol{\chi}$, $\mathbf{\Pi}_{1} \leftarrow \boldsymbol{\omega}_1$, and $\mathbf{\Pi}_{2} \leftarrow \boldsymbol{\omega}_2$ for SAC algorithm. In the meta-adaptation stage, the agent randomly samples a batch of another replay buffer denoted by $\mathcal{B}^{\mathrm{ada}}$ and updates the AN and CN in the modified-DDPG algorithm, and AN, CN1, and CN2 in the SAC algorithm as follows
\begin{align}\label{eq21}
\left\{ \begin{array}{l}
\mathbf{\Delta} = \mathbf{\Delta}-\bar\epsilon_{1}{\nabla _{\Delta} }{{\mathcal{L}}({\mathbf{\Delta}}, \mathcal{B}^{\mathrm{ada}})},\\
\mathbf{\Omega}= \mathbf{\Omega}-\bar\epsilon_{2}{\nabla _{\Omega} }{{\mathcal{L}}({\mathbf{\Omega}}, \mathcal{B}^{\mathrm{ada}})},
\end{array} \right.
\end{align}
where $\bar\epsilon_{1}$ and $\bar\epsilon_{2}$ denote the learning rates of the modified-DDPG.
\small
\begin{align}\label{eq55}
\left\{ \begin{array}{l}
\mathbf{\Xi} = \mathbf{\Xi}-\bar\epsilon_{3}{\nabla _{\Xi}}{{\mathcal{L}}({\mathbf{\Xi}},  \mathcal{B}^{\mathrm{ada}})},\\
\mathbf{\Pi}_{1}= \mathbf{\Pi}_{1}-\bar\epsilon_{4}{\nabla _{\Pi_{1}} }{{\mathcal{L}}({\mathbf{\Pi}_{1}}, \mathcal{B}^{\mathrm{ada}})},\\
\mathbf{\Pi}_{2}= \mathbf{\Pi}_{2}-\bar\epsilon_{5}{\nabla _{\Pi_{2}} }{{\mathcal{L}}({\mathbf{\Pi}_{2}}, \mathcal{B}^{\mathrm{ada}})},
\end{array} \right.
\end{align}
\normalsize
where $\bar\epsilon_{3}$, $\bar\epsilon_{4}$, and $\bar\epsilon_{5}$ represent the learning rates of the SAC algorithm. Algorithm \ref{meta-adaptation-phase} is referred to as proposed Scheme 2, which is a combination of Algorithm \ref{Algorithm 1} and the MMDS algorithm. The Algorithm \ref{meta-training-phase} actually serves as a pre-training algorithm, adjusting the weights of neural networks in the modified-DDPG and SAC algorithms. As explained before, the weights of neural networks used in the meta-adaptation phase are initialized with the weights adjust in the meta-training phase. Essentially, Algorithm \ref{meta-adaptation-phase} is our proposed algorithm, which we identify as "proposed Scheme $2$". 
\vspace*{-0.1em}
\section{Simulation \ Results}\label{simulation}
	In this section, we assess the effectiveness of the proposed resource allocation framework for optimising the discussed active STAR-RIS-aided SWIPT system through extensive simulation results.  To this goal, we initially present our simulation configuration and initialize the parameters. Subsequently, we analyze the performance of the system when varying key system parameters.
	
	\vspace*{-1.3em}
	\subsection{Simulation Configuration} 
	A three-dimensional (3D) coordinate system is considered, where the BS and the active STAR-RIS are located at coordinates $(0, 0, 15)$~m and $(0, 20, 15)$~m, respectively. Moreover, it is assumed that users in the reflection  area of the STAR-RIS are uniformly distributed within a circle centered at $(0, 16, 0)$~m with a radius of $r_{r}=3$~m. Similarly, users in the re-transmission area of the STAR-RIS are located within a circle centered at $(0, 24, 0)$~m  with a radius of $r_{t}=3$~m based on a uniform random distribution.
	The channel models between the BS and users in the reflection and re-transmission area capture LoS links only and thus follow the Rayleigh fading model. 
	Moreover, the channel gains between the active STAR-RIS and users in the reflection and re-transmission areas and also the channel gains between the BS and active STAR-RIS are considered to capture both LoS and NLoS links and thus follow the Rician fading with a Rician factor of $5$dB. 

\normalsize

\vspace*{2em}
\begin{table}[t!] 
	\vspace{-1 em}
	\centering
	\caption{\color{black}{Simulation Parameters}}
	\label{tab:my_label}
	\begin{tabular}{ |l|l||l|l| } 
		\hline
		\textbf{Parameter} & \textbf{Value} & \textbf{Parameter} & \textbf{Value} \\
		\hline
		\hline		
		\hline 
		$P_{\textrm{BS}}^{\max}$ & $40$~dBm &
        $\sigma^{2}_{\textrm{t}, j}$, $\sigma^{2}_{\textrm{r}, z}$ & $-80$~dBm \\
		\hline $P_{\textrm{I}}^{\max}$ & $5$~dBm &
		$M_{\textrm{r}, i}, M_{\textrm{t}, j}$ & $0.02$~Watt \\
		\hline $a_{\textrm{r}, i}, a_{\textrm{t}, j}$ & 6400 &
		$b_{\textrm{r}, i}, b_{\textrm{t}, j}$ & 0.003 \\
		\hline $\sigma^{2}_{\textrm{r}, i}, \sigma^{2}_{\textrm{r}, z}$,  & $-8
		0$~dBm &
		$\delta^{2}_{\textrm{r}, i}, \delta^{2}_{\textrm{t}, j} $ & $-50$~dBm \\
		\hline $\mu_{\textrm{r}, i}, \mu_{\textrm{t}, j}$ & $0.5$ &
		${P_{\textrm{BS}}^{\textrm{Cir}}}$ & $30$~dBm \\
		\hline $p_{i}^{\textrm{Cir}}, p_{j}^{\textrm{Cir}}$ & $7$~dBm &
		$P_{c}$ & $-10$~dBm \\
		\hline $P_{\textrm{DC}}$ & $-5$~dBm &
		$\zeta \triangleq \nu^{-1}$ & 1.25 \\
		\hline $N_{T}$ & 5 &
		$U_{\textrm{r}} = U_{\textrm{t}}$ & 3 \\
		\hline $M$ & 16 & $K_{\max}$ & $10$ \\
		\hline 
	\end{tabular}
	\label{tab2}
	\vspace{-0.75 em}
\end{table}
	\begin{figure}
	\centering
	\includegraphics[scale=0.3]{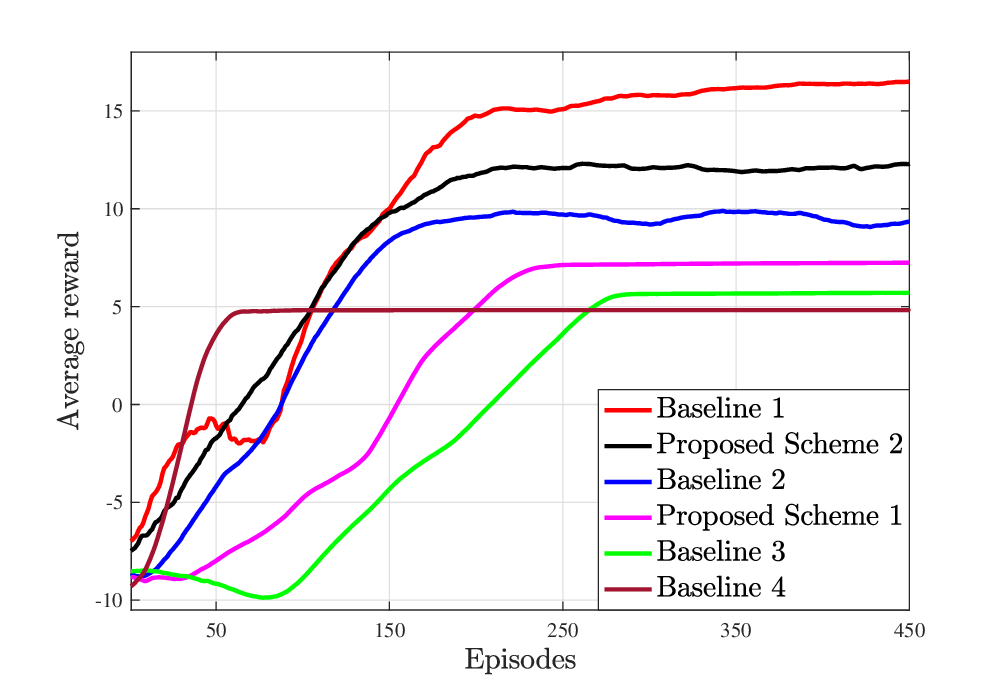}
	\caption{Average reward vs. the number of episodes.}
	\label{convergence}
\end{figure}
\vspace*{-3em}
\subsection{System Utility}
To assess the performance of the proposed resource allocation scheme on the discussed system model, we utilize average reward and average system EE as the key performance metrics. In our evaluation, we consider the following benchmark schemes:
\begin{enumerate}
	\item Baseline 1: This baseline adopts Algorithm \ref{meta-adaptation-phase} for resource allocation and evaluates the performance of the active STAR-RIS-aided SWIPT system with linear EH model.
	\item proposed Scheme 2: This baseline is similar to \text{Baseline 1}, except for adopting a non-linear EH model for SWIPT.
	\item Baseline 2: This baseline is also similar to \text{Baseline 1}, except for considering $\lbrace \rho_{\textrm{r}, i} \rbrace_{i \in \mathcal{U}_{\textrm{r}}}, \lbrace \rho_{\textrm{t}, j} \rbrace_{j \in \mathcal{U}_{\textrm{t}}} = 0.5$.
	\item Proposed Scheme 1: This baseline adopts Algorithm \ref{prposed schem 1} for resource allocation.
	\item Baseline 3: This baseline adopts Algorithm \ref{meta-adaptation-phase}. However, the phase shifts for reflection and re-transmission $\mathbf{\Theta}_{\textrm{r}}$ and $\mathbf{\Theta}_{\textrm{t}}$ are randomized.
	\item Baseline 4: This baseline adopts Algorithm \ref{meta-adaptation-phase} for resource allocation, however by considering a passive STAR-RIS-aided SWIPT system.
\end{enumerate}
\par Fig.~\ref{convergence} illustrates the convergence behavior of the different baselines we introduced earlier. We observe that Baseline $1$ outperforms the others, especially the proposed Scheme $2$. This observation highlights that the linear impractical EH model incorporated in this baseline, provides a benchmarking use-case only. In contrast, proposed Scheme $2$ achieves a lower average reward, due to incorporating the practical non-linear EH model. In addition, this baseline outperforms the others, thanks to better adaptability and generalization that its incorporated meta-learning feature offers. At the cost of lower achieved average reward, Baseline 2 converges a bit faster than Baseline $1$ and proposed Scheme $2$, since this baselines does not optimise the PS ratios $\lbrace \rho_{\textrm{r}, i} \rbrace_{i \in \mathcal{U}_{\textrm{r}}}$ and $\lbrace \rho_{\textrm{t}, j} \rbrace_{j \in \mathcal{U}_{\textrm{t}}}$, thus enjoys a smaller action space. We also observe that the proposed Scheme $1$ has lower achieved average reward, compared to prior baselines we discussed. This is due to lacking meta-learning feature, which limits its adaptability and therefore this baseline experiences a longer time to converge. In spite of adopting random phase shifts for the active STAR-RIS and also a more limited action space, the Baseline $3$ exhibits a longer convergence behaviour in comparison with all other baselines. This is due to the lack of optimal selection of the phase shift matrices $\mathbf{\Theta}_{\textrm{r}}$ and $\mathbf{\Theta}_{\textrm{t}}$. In addition, the randomized phase shifts of active STAR-RIS provide the system with much lower average reward value achieved for this baseline. Eventually, Baseline 4 has the worst performance from the average reward perspective, compared to prior baselines. This is due to lack of a proper amplification mechanism. However, the fastest convergence among baselines is achieved by this baselines, thanks to its limited action space.

\begin{figure}
	\centering
	\includegraphics[scale=0.3]{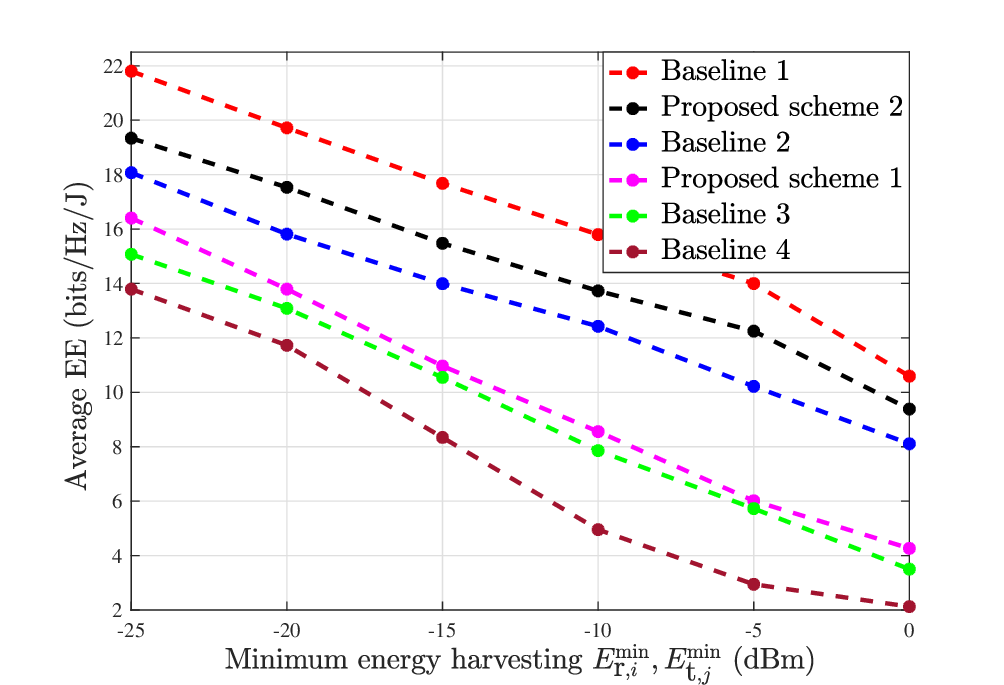}
	\caption{Average EE vs. minimum harvested power.}
	\label{fig:a}
\end{figure}
\vspace*{-1em}
\begin{figure}
	\centering
	\includegraphics[scale=0.3]{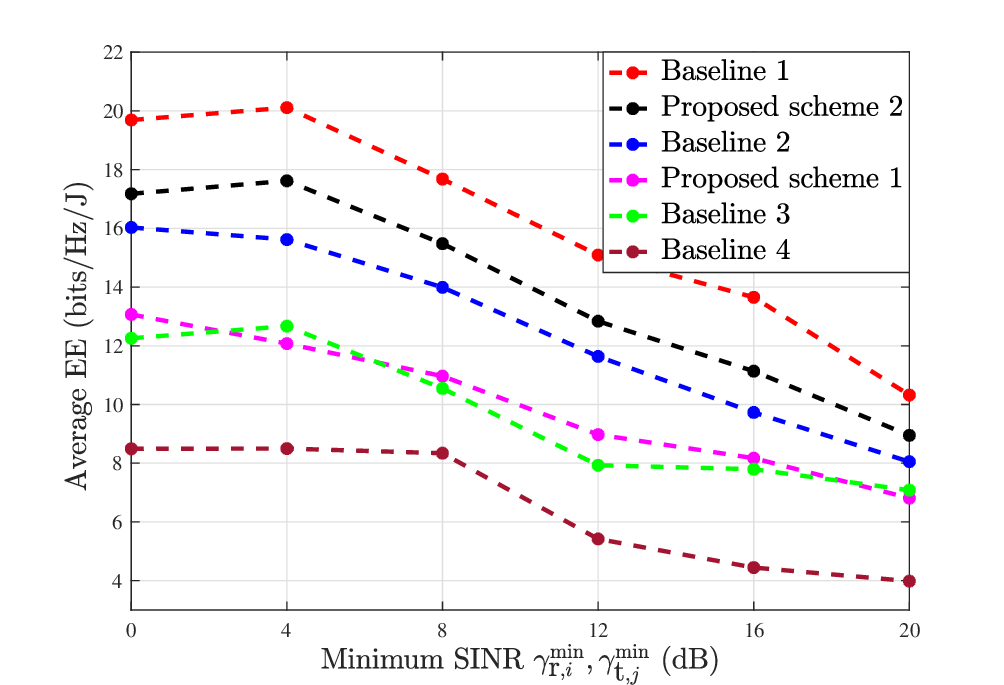}
	\caption{Average EE vs. minimum SINR.}
	\label{fig:b}
\end{figure}

\vspace*{1em}
Fig.~\ref{fig:a} depicts the impact of the minimum EH requirement of users on average system EE. Evidently, by increasing the minimum EH requirement of users, a diminishing trend is observed for the average system EE. This observation stems from the fact that ensuring a higher minimum EH requirement for each user, as dictated by constraints $\text{C}_3$ and $\text{C}_4$ in \eqref{eq5}, narrows down the solution space of the optimisation problem $\mathcal{P}$. Consequently, this constraint tightening degrades the value of objective function in $\mathcal{P}$. In addition, we observe that the Baseline 1 achieves the most average system EE, however, under the ideal and impractical assumption of linear EH model for SWIPT. In comparison, the proposed Scheme 2 adopts the practical non-linear EH model at the cost of relatively lower average system EE. Compared with Baseline 2, the superiority of the Baseline 1 and the proposed Scheme 2 signifies the importance of optimising the PS ratio $\lbrace \rho_{\textrm{r}, i} \rbrace_{i \in \mathcal{U}_{\textrm{r}}}$ and $\lbrace \rho_{\textrm{t}, j} \rbrace_{j \in \mathcal{U}_{\textrm{t}}}$ of SWIPT technology. We also observe a similar performance for the proposed Scheme 1 and the Baseline 3, which respectively highlight the importance of meta-learning, as well as optimising the phase shifts of the active STAR-RIS, i.e., $\mathbf{\Theta}_{\textrm{r}}$ and $\mathbf{\Theta}_{\textrm{t}}$. Lastly, the worst performance definitely corresponds to the Baseline 4, which adopts the conventional passive model for the STAR-RIS~\cite{passive_RIS}. 

By maintaining the same order of baselines of Fig.~\ref{fig:b}, the effect of the target SINR requirement of users i.e., $\gamma^{\min}_{\textrm{r}, i}$ and $\gamma^{\min}_{\textrm{t}, j}$ on average system EE is illustrated in Fig.~\ref{fig:b}. It is evident that all baselines display a declining trend as the target SINR requirements increase. This trend is attributed to the fact that enforcing higher target SINR requirements for each user, as mandated by constraints $\text{C}1$ and $\text{C}2$ in \eqref{eq5}, results in increased transmit power consumption by the BS to meet the user demands. At lower values of $\gamma^{\min}_{\textrm{r}, i}$ and $\gamma^{\min}_{\textrm{t}, j}$ on the x-axis, the baselines exhibit a gradual decrease or even remain relatively stable, whereas at higher values on the x-axis, the trend becomes more pronounced. This indicates that the impact of power consumption on average system EE becomes more significant than the impact of data rate.

\begin{figure}
	\centering
	\includegraphics[scale=0.3]{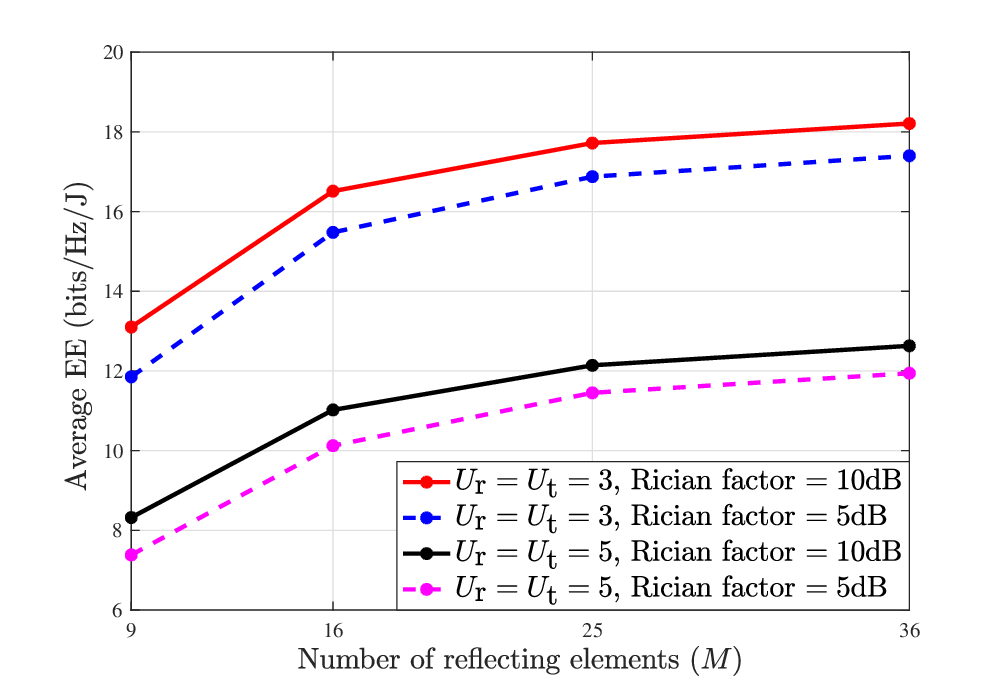}
	\caption{Average EE vs. minimum harvested power.}
	\label{reflecting_elements}
\end{figure}
\vspace*{-1.5em}
\begin{figure}
	\centering
	\includegraphics[scale=0.3]{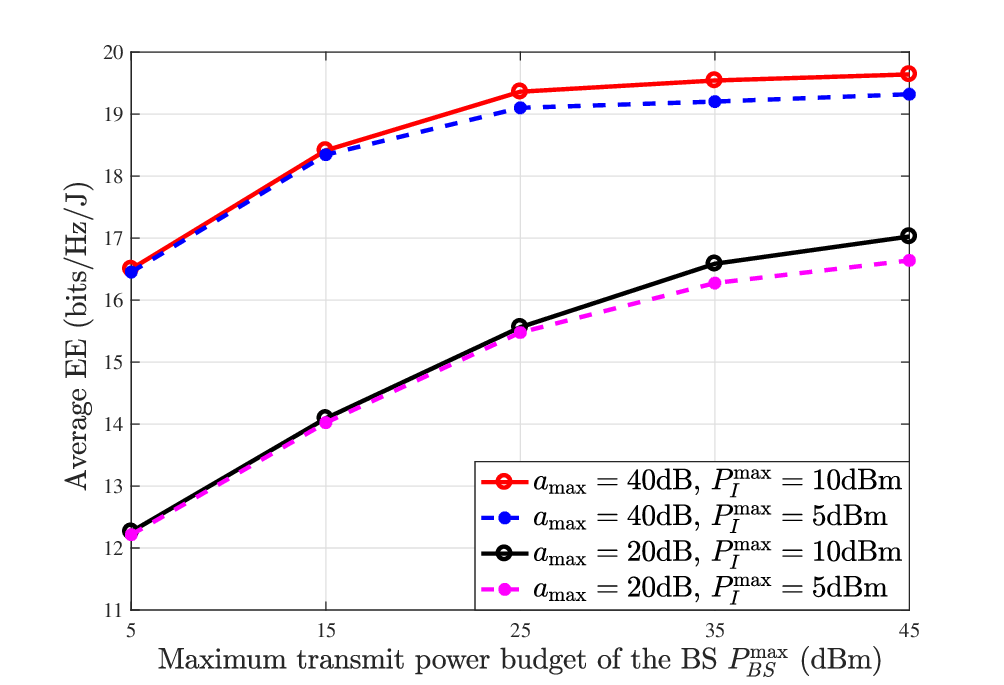}
	\caption{Average EE vs. minimum SINR.}
	\label{power_BS}
\end{figure}

\vspace*{2em}
In Fig.~\ref{reflecting_elements}, we analyze the average system EE, by varying the number of active STAR-RIS reflecting elements in x-axis. The baselines include different number of users and values of the Rician factor. First, by enlarging the active STAR-RIS, the average system EE exhibits an increasing trend, due to better covering the wireless environment. In addition, the higher number of users served in this system, the lower average system EE is achieved. This observation stems from the fact that incorporating more users leads to more mutual interference, which thereby lowers the achievable aggregate data rate. Moreover, the BS requires more transmit power to support a higher number of users. One can also find that the higher Rician factor is, the higher average system EE is achieved. This is because a larger value of Racian factor provides the system with a stronger LoS path for wireless signal propagation.

\par Fig.~\ref{power_BS} specifies the variations of system EE in the y-axis for different values of $a_{\text{max}}$ and $P_{I}^{\text{max}}$ in baselines, and $P^{\text{max}}_{\text{BS}}$ in the x-axis. As $P^{\text{max}}_{\text{BS}}$  increases, the system EE improves for all baselines, albeit in a saturating fashion. This behaviour arises because at lower values of $P^{\text{max}}_{\text{BS}}$, the system data rate dominates the system power consumption, whereas the opposite holds at higher values of $P^{\text{max}}_{\text{BS}}$. In addition, we find that for a given $P^{\text{max}}_{{I}}$, larger values of $a_{\text{max}}$ allow for a wider range of the amplification factor at the active STAR-RIS for strengthening the incident signals. Consequently, better system data rates and EE are expected. Meanwhile, for a given value of $a_{\text{max}}$, a larger amplification power budget $P^{\text{max}}_{{I}}$, at the active STAR-RIS results in superior system EE. This is because the active STAR-RIS can offer more amplification gain while utilizing the same transmit power of the BS. Therefore, dynamic adjustment of the active STAR-RIS's amplification gain in accordance with the BS's transmit power is crucial in the design of this system. Another notable observation drawn from this figure corresponds to the gap between the baselines. Particularly, for smaller values of $P^{\text{max}}_{\text{BS}}$ and a given $a_{\text{max}}$, a couple of dash and solid lines grow tightly close, whereas for larger values in x-axis, the gap widens. Comparatively, given a value of $P^{\text{max}}_{I}$, the gap between two solid lines or two dash lines is significant. This highlights the relatively minor impact of  $P^{\text{max}}_{\text{I}}$ on system EE, in comparison with the influence of $a_{\text{max}}$.
\vspace*{-0.5em}
\section{Conclusion}\label{Conclusion}
In this paper, we studied an energy efficient active STAR-RIS-aided SWIPT system. To evaluate the performance of this system, we formulate a system EE maximization problem by jointly optimising the amplification factor, element selection and phase shift matrices of the active STAR-RIS, the transmit beamforming of the BS, as well as the PS ratio.  
We proposed a real-time and adaptive solution policy based on meta-reinforcement learning and convex optimisation techniques. According to the simulations, the proposed active STAR-RIS-assisted SWIPT system outperforms its passive counterpart by 57\% system EE gain in average.

\bibliographystyle{IEEEtran}
\bibliography{myref}
\end{document}